\RequirePackage{fix-cm}
\documentclass[twocolumn, natbib]{svjour3}          % twocolumn
\smartqed  % flush right qed marks, e.g. at end of proof
\usepackage{times}
\usepackage{epsfig}
\usepackage{graphicx}
\usepackage[cmex10]{amsmath}
\usepackage{amssymb}
\usepackage{tabulary,color}
\usepackage{tabularx}
\usepackage{multirow}
\usepackage{booktabs, siunitx, dcolumn}
\usepackage{colortbl}
\usepackage{pifont}
\usepackage{dsfont}
\usepackage{bbding}
\usepackage{enumitem} % noindent for itemize
\usepackage[para]{threeparttable} % footnote for table
\usepackage[export]{adjustbox}
\usepackage[percent]{overpic}
\usepackage{array} % ALIGNMENT PACKAGES
\usepackage{url} % PDF, URL AND HYPERLINK PACKAGES
\usepackage{fixltx2e} % FLOAT PACKAGES
\usepackage{algorithm}     % algorithm
\usepackage{algorithmic}     % algorithm
 \usepackage[caption=false,font=footnotesize]{subfig}
\usepackage{natbib}
\setcitestyle{authoryear,open={(},close={)}} % cite style

\usepackage[pagebackref=true,breaklinks=true,colorlinks,citecolor=blue,bookmarks=false]{hyperref} % cite color

\usepackage[toc,page]{appendix} % appendix

\journalname{International Journal of Computer Vision}

\begin{document}\sloppy

\title{Memory-augmented Deep Unfolding Network for Guided Image Super-resolution}

\author{Man Zhou$^{1,2}$, \; Keyu Yan$^{1,2}$, \; Jinshan Pan$^{3}$, \; Wenqi Ren$^{4}$, \; Qi Xie$^{5}$, \; Xiangyong Cao$^{5}$}

\authorrunning{International Journal of Computer Vision}
\institute{
 Corresponding author: Xiangyong Cao \at
           \email{caoxiangyong@mail.xjtu.edu.cn} 
		   \and
		   Man Zhou and Keyu Yan contributed equally \at
           \email{manman@mail.ustc.edu.cn}
           \and
	$^1$\;\; Hefei Institute of Physical Science, Chinese Academy of Sciences, Hefei, China \\
	$^2$\;\; University of Science and Technology of China, Hefei, China \\
	$^3$\;\; Nanjing University of Science and Technology, Nanjing, China \\
	$^4$\;\; Institute of Information Engineering, Chinese Academy of Sciences, Beijing, China \\
	$^5$\;\; Xi’an Jiaotong University, Xi’an, China
}
% The e-mail address, and telephone number(s) of the corresponding author
\date{}

\maketitle

\begin{abstract}
Guided image super-resolution (GISR) aims to obtain a high-resolution (HR) target image by enhancing the spatial resolution of a low-resolution (LR) target image under the guidance of a HR image. However, previous model-based methods mainly takes the entire image as a whole, and assume the prior distribution between the HR target image and the HR guidance image, simply ignoring many non-local common characteristics between them. To alleviate this issue, we firstly propose a maximal a posterior (MAP) estimation model for GISR with two types of prior on the HR target image, i.e., local implicit prior and global implicit prior. The local implicit prior aims to model the complex relationship between the HR target image and the HR guidance image from a local perspective, and the global implicit prior considers the non-local auto-regression property between the two images from a global perspective. Secondly, we design a novel alternating optimization algorithm to solve this model for GISR.  The algorithm is in a concise framework that facilitates to be replicated into commonly used deep network structures. Thirdly, to reduce the information loss across iterative stages, the persistent memory mechanism is introduced to augment the information representation by exploiting the Long short-term memory unit (LSTM) in the image and feature spaces. In this way, a deep network with certain interpretation and high representation ability is built. Extensive experimental results validate the superiority of our method on a variety of GISR tasks, including Pan-sharpening, depth image super-resolution, and MR image super-resolution.
 \end{abstract}

% provide an abstract of 150 to 250 words. (177 words)
% provide 4 to 6 keywords ( 6 keywords )

\begin{figure*}
\centering
{\includegraphics[width = 7in]{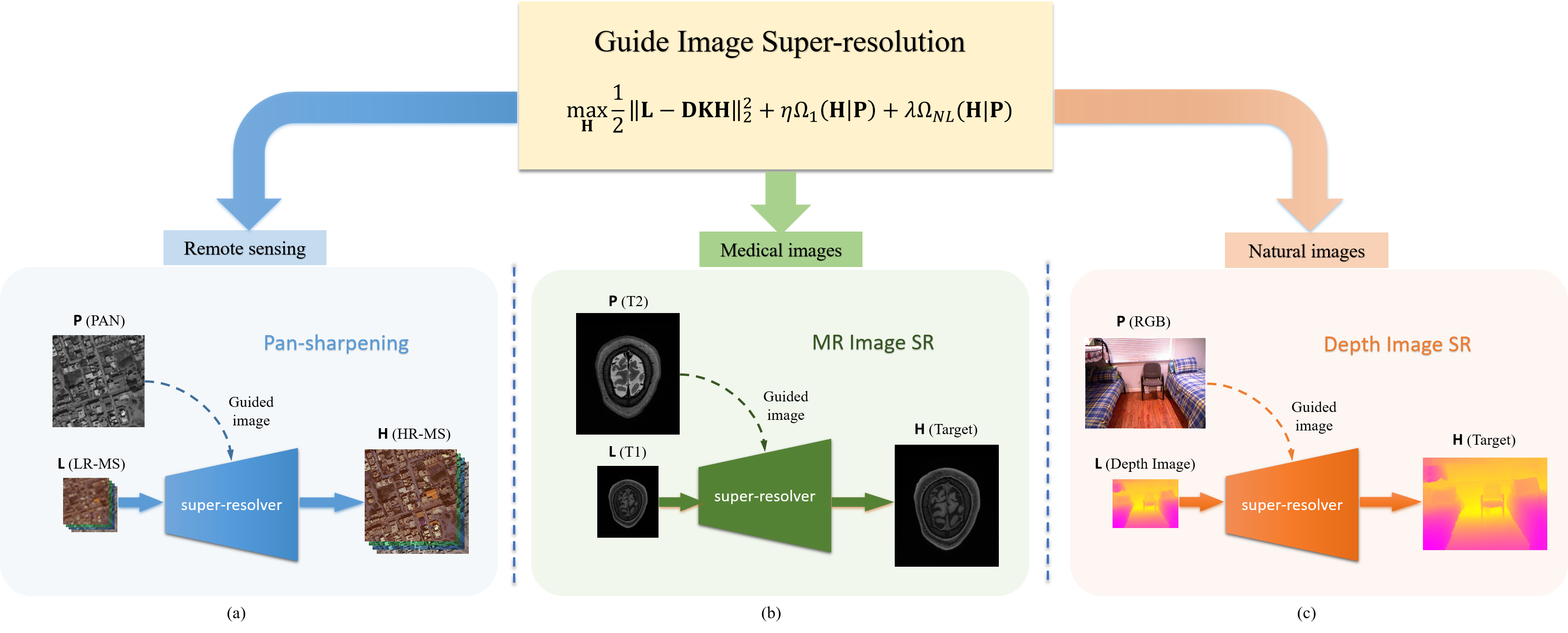}}
\caption{Typical Guided Image Super-Resolution (GISR) tasks. (a) Pan-sharpening. (b) MR Image SR. (c) Depth Image SR.}
\label{fig.abs}
\end{figure*}

\section{Introduction}\label{sec:introduction}
Guided image super-resolution (GISR) has attracted substantial attention and achieved remarkable progress in recent years. Different from the single image super-resolution (SISR) task that only receives input with a single low-resolution (LR) target image, GISR aims to super-resolve the LR target image under the guidance of an additional high-resolution (HR) image whose structural details can help enhance the spatial resolution of the LR target image. Additionally, the guided image in GISR is capable of regularizing the super-resolution process and alleviating the ill-posed problem in SISR, thus leading to better performance. 

In GISR, the degradation model is assumed to be the same with SISR, and can be mathematically formulated as
\begin{equation}\label{eq:ab}
     \mathbf{L} = (\mathbf{H}\ast \mathbf{k})\downarrow_{\mathbf{s}} + \mathbf{n_s}  
\end{equation}
where $\ast$ represents the convolution operation, $\mathbf{L}$ denotes the LR target image, which is obtained by applying the blurring kernel $\mathbf{k}$ and down-sampling operation $\downarrow_{\mathbf{s}}$ on the HR target image $\mathbf{H}$, and $\mathbf{n_s}$ is usually assumed to be additive white Gaussian noise (AWGN) with standard deviation $\sigma$. Different from SISR, GISR uses additional image to guide the super-resolution process. Typical GISR tasks include pan-sharpening~\citep{masi2016pansharpening,Xu_2021_CVPR,cao2021pancsc}, depth image super-resolution~\citep{DKN,PacNet,CTKT}, and magnetic resonance (MR) image super-resolution~\citep{oktay2016multi,pham2017brain,wang2020enhanced}, which are illustrated in Fig.~\ref{fig.abs}. Specifically, pan-sharpening can be considered as a PAN-guided multi-spectral (MS) image SR task, where MS images suffer from the low spatial resolutions issue due to the physical hardware limits, and the texture-rich HR PAN images captured from the same scene can act as the guidance image to enhance the spatial resolution of LR MS image. For depth image super-resolution, the HR color image are exploited as a prior for reconstructing regions in depth image, which contains semantically-related and structure-consistent content with the color image. The LR depth image and the HR color image characterize the target and guidance image, respectively. For MR image super-resolution, it aims to super-resolving the MR image of a target contrast under the guidance of the corresponding auxiliary contrast, which provides additional anatomical information. The MR image and the corresponding auxiliary contrast image characterize the target and guidance image, respectively.

Recently, researchers have proposed a large number of guided image super-resolution (GISR) approaches. The main idea of these approaches is transferring structural details of the guidance image to the target image. A large class of methods is filter-based, such as bilateral filtering~\citep{tomasi1998bilateral,kopf2007joint}, and guided image filtering~\citep{he2012guided,shen2015multispectral,ham2017robust}, which firstly learns the weights of filter from the guidance image and then applies the learnt weights to filter the target image. These approaches are implemented by hand-crafted objective functions, and may not reflect image priors well. The issue can be alleviated by model-based methods, such as sparsity-based models~\citep{deng2019deep,deng2020deep} and coupled dictionary learning-based models~\citep{wang2012semi,zhuang2013supervised,liu2014semi,jing2015super,bahrampour2015multimodal}. These model-based methods aim to capture the correlation among different image modalities by placing explicit hand-crafted priors on different image modalities. Although this type of methods have shown good performance, their priors require careful design and they often require  high-computational cost for optimization, limiting their practical applications. Furthermore, the limited representation ability of handcrafted priors leads to unsatisfactory results when processing complex scenes.

Inspired by the success of deep learning in other computer vision tasks, multi-modal deep learning approaches have been developed ~\citep{ngiam2011multimodal,li2016deep,wu2018fast}. A common method is to use a shared layer for fusing different input modalities~\citep{ngiam2011multimodal}. Following this idea,  a deep neural network (DNN) based joint image filtering method which utilizes the structures of both target and guidance images, was put forward in \citep{li2016deep}. A deep neural network (DNN) reformulation of guided image filtering was designed in \citep{wu2018fast}. However, these DNNs are black-box models in the sense that they neglect the signal structure and properties of the correlation across modalities. To alleviate this black-box issue, some attempts have been made based on the model-driven deep learning methodology. \citep{deng2019deep} proposed a deep unfolding network using the deep unfolding design LISTA~\citep{gregor2010learning} that solves the sparse coding model. \citep{marivani2020multimodal,deng2020deep} proposed multi-modal deep unfolding networks that perform steps similar to an iterative algorithm for convolutional sparse coding model. Nevertheless, current deep unfolding networks~\citep{deng2019deep,marivani2020multimodal,deng2020deep} are all built on the (convolutional) spare coding models, where the (convolutional) sparse codes across modalities are assumed to be close. Therefore, these methods only consider some hand-crafted prior and thus do not fully explore the correlation distribution of different modal images. Additionally, the potential of cross-stages for the deep unfolding network has not been fully explored since feature transformation between adjacent stages with reduced channel number leads to information loss. In short, current state-of-the-art deep unfolding  methods for GISR suffer from two issues: 1) the correlation distribution of different modal images is not fully exploited, and 2) the unfolding network suffers from severe information loss in the signal flow. To address the aforementioned issues, this paper first proposes a general GISR model by fully considering the correlation distribution across modalities, and then designs an interpretable information-persistent deep unfolding network by exploring the short-and-long term memory mechanisms. 

Specifically,  we firstly construct a universal variational model for the GISR problem from the maximum a posterior (MAP) perspective by simultaneously considering two well-established data priors, i.e.,  local implicit prior and global implicit prior. The local implicit prior models the relationship between the HR target image and the HR guidance image implicitly from a local perspective, and thus can help capture the local related information between the target image and the guidance image. The global implicit prior considers the non-local auto-regression property between the two images from a global perspective, and thus the global correlation between the two images can be well exploited. Since the scene of the target and guidance image in the GISR problem is almost the same, both images thus contains repetitively similar patterns, matching the motivation of the designed non-local auto-regression prior. Secondly, we design a novel alternating optimization algorithm for the proposed variational model, and then unfold the algorithm into a deep network in an effective and transparent manner with cascaded multi-stages. Each stage in the implementation corresponds to three interconnected sub-problems, and each module connects with a specific operator of the iterative algorithm. The detailed flowchart is illustrated in Fig.~\ref{mainfig} and the sub-problems are remarked by different colors. In addition, to facilitate the signal flow across iterative stages, the persistent memory mechanism is introduced to augment the information representation by exploiting the long-short term unit in the image and feature spaces. In the three sub-problems, the output feature maps of each iterative module are selected and integrated for the next iterative stage, thus promoting information fusion across stages and reducing the information loss. In this way, both the interpretation and representation ability of the deep network can be improved. Extensive experimental results validate the superiority of the proposed algorithm against other state-of-the-art methods over the three typical GISR tasks, i.e., pan-sharpening, depth image super-resolution, MR image super-resolution.

In summary, our contributions are four-fold:
\begin{itemize}
  \item We propose a new GISR model by embedding two image priors, i.e., local implicit prior and global implicit prior, from a maximum a posterior (MAP) perspective. The two-prior model framework and carefully designed algorithm allows us to separately adopt different type of network modulars for the two type of implicit prior when constructing GISR network by adopting deep unrolling technique. Then, we design a novel alternating optimization algorithm for the proposed model.
  \vspace{0.4em}

  \item We propose an interpretable memory-augmented deep unfolding network (MADUNet) for the GISR problem by unfolding the iterative algorithm into a multistage implementation, which incorporates the advantages of both the model-based prior-equipped methods and data-driven deep-learning methods. With such design, the interpretation of the deep model is improved.  
  \vspace{0.4em}
  
  \item We propose a new memory mechanism and design a non-local cross-modality module to alleviate the severe information loss issue in the signal flow. The former selectively integrates the features of different-layer and intermediate output of previous stages into the next stage, while the latter explores the information interaction in the target images and across two modalities of the target and guidance images. With such design, the representation ability of the deep model is improved.  
  \vspace{0.4em}
  
  \item Extensive experiments over three representative GISR tasks, i.e., pan-sharpening, depth image super-resolution and MR image super-resolution demonstrate that our proposed network outperforms other state-of-the-art methods both qualitatively and quantitatively.
\end{itemize}

\vspace{-.4cm}
\section{Related Work}
\vspace{-.2cm}
\subsection{Single Image Super-resolution}
Single image super-resolution (SISR) methods have been extensively studied in recent decades, and these methods can be roughly divided into three categories, i.e., interpolation-based approaches, reconstruction-based approaches, and learning-based approaches. 

Interpolation-based methods~\citep{dai2007soft,sun2008image,sanchez2008noniterative} is easy to implement, but tend to overly smooth image edges and bring in aliasing and blurring effects. Reconstruction-based methods~\citep{mallat2010super,dong2012nonlocally,yang2013fast} utilize various image prior to regularize the ill-posed SISR problem, and always outperform the interpolation-based methods. Learning-based methods~\citep{yang2008image,yang2010image,yang2012coupled,jia2012image,timofte2013anchored,timofte2014a+,dong2015image,kim2016accurate,kim2016deeply,mao2016image,bruna2015super,zhang2018image,tai2017image,zhang2018residual} is implemented by directly learning the mapping function from LR image to HR image using some machine learning techniques. For example, several sparse representation and dictionary learning based methods were proposed in~\citep{yang2008image,yang2010image,yang2012coupled,jia2012image}, which assume that HR/LR image pairs share the same sparse coefficients in terms of a pair of HR and LR dictionaries. Later, several anchored neighbourhood regression based methods were developed in~\citep{timofte2013anchored,timofte2014a+}, which combines the neighbor embedding technique and dictionary learning method. Recently, deep learning methods have become the most popular technique among the learning-based methods for SISR. SRCNN~\citep{dong2015image} was the first deep convolutional network for SISR. A faster version FSRCNN was designed in~\citep{dong2016accelerating}. Subsequently, a series of SISR deep learning methods focus on improving the network complexity by increasing the depth of the network. For example, a 20-layer residual network for SR was proposed in~\citep{kim2016accurate,kim2016deeply}, a 30-layer convolutional autoencoder was developed in~\citep{mao2016image}, and a 52-layer deep recursive residual network was designed in~\citep{tai2017image}. Additionally, residual learning has been used to learn inter-layer dependencies in~\citep{zhang2018image,zhang2018residual}. Deep unfolding technique has also been applied to SISR in~\citep{liu2016robust,zhang2020deep}, where the iterative algorithm is implemented by a neural network.

\vspace{-.6cm}
\subsection{Guided Image Super-resolution}
\vspace{-.2cm}
Compared with single image SR, guided image SR utilizes an additional HR guidance image to super-resolve the LR target image by transferring structural information of the guidance image to the target image.

The bilateral filter~\citep{tomasi1998bilateral} is essentially a translation-variant edge-preserving filter, which calculates a pixel value by the weighted average of its neighboring pixels, and the weights are learnt from the given image using filter kernels. The joint bilateral upsampling~\citep{kopf2007joint} is a generalization of the bilateral filter. Instead of computing the weights from the input image, the joint bilateral upsampling learns the weight from an additional guidance image using a range filter kernel. In this way, the high frequency components of the guidance image can be transfered into the LR target image. 
However, this method may introduce gradient reversal artifacts. To mitigate this issue, guided image filtering~\citep{he2012guided} was designed by directly transferring the guidance gradients into the LR target image. But this method still bring in notable appearance change. To alleviate this issue, \citep{shen2015multispectral} put forward a model to optimize a novel scale map for capturing the nature of structure discrepancy between the target and guidance images. However, since the above methods only consider  the static guidance image, these methods may convert inconsistent structure and detail information of guidance image to the target image. \citep{ham2017robust} developed a robust guided image filtering method to iteratively refine the target image by utilizing the static guidance image and the dynamic target image. \citep{song2019multimodal} proposed a coupled dictionary learning based GISR method, which learns a set of dictionaries that couple different image modalities in the sparse feature domain from a training dataset, and thus alleviate the inconsistency issue between the guidance and target images. These approaches adopt hand-crafted objective functions, and may not reflect natural image priors well. 

Recently, deep learning-based methods have been applied for this problem. A deep neural network (DNN) based joint image filtering method which simutaneously utlizes the structures of both target and guidance images, was put forward in \citep{li2016deep}. A deep neural network (DNN) reformulation of guided image filtering was designed in \citep{wu2018fast}. The two methods are both pure data-driven methods and rely on large amount of image pairs to train the network. Additionally, based on the model-driven deep learning methodology, \citep{marivani2020multimodal} proposed a multimodal deep unfolding network that performs steps similar to an iterative algorithm for convolutional sparse coding with side information. \citep{deng2019deep} proposed another deep unfolding network for guided image SR using the deep unfolding design LISTA~\citep{gregor2010learning}, where the latent representations of the input images are computed by two LISTA branches, and the HR target image is finally generated by the linear combination of these representations. Later, similar method was proposed in~\citep{deng2020deep} by using three LISTA branches to separate the common feature shared by different modalities and unique features for each modality, and the target image is reconstructed by the combination of these common and unique feature.

\begin{figure*}
\begin{center}
\includegraphics[width=\textwidth]{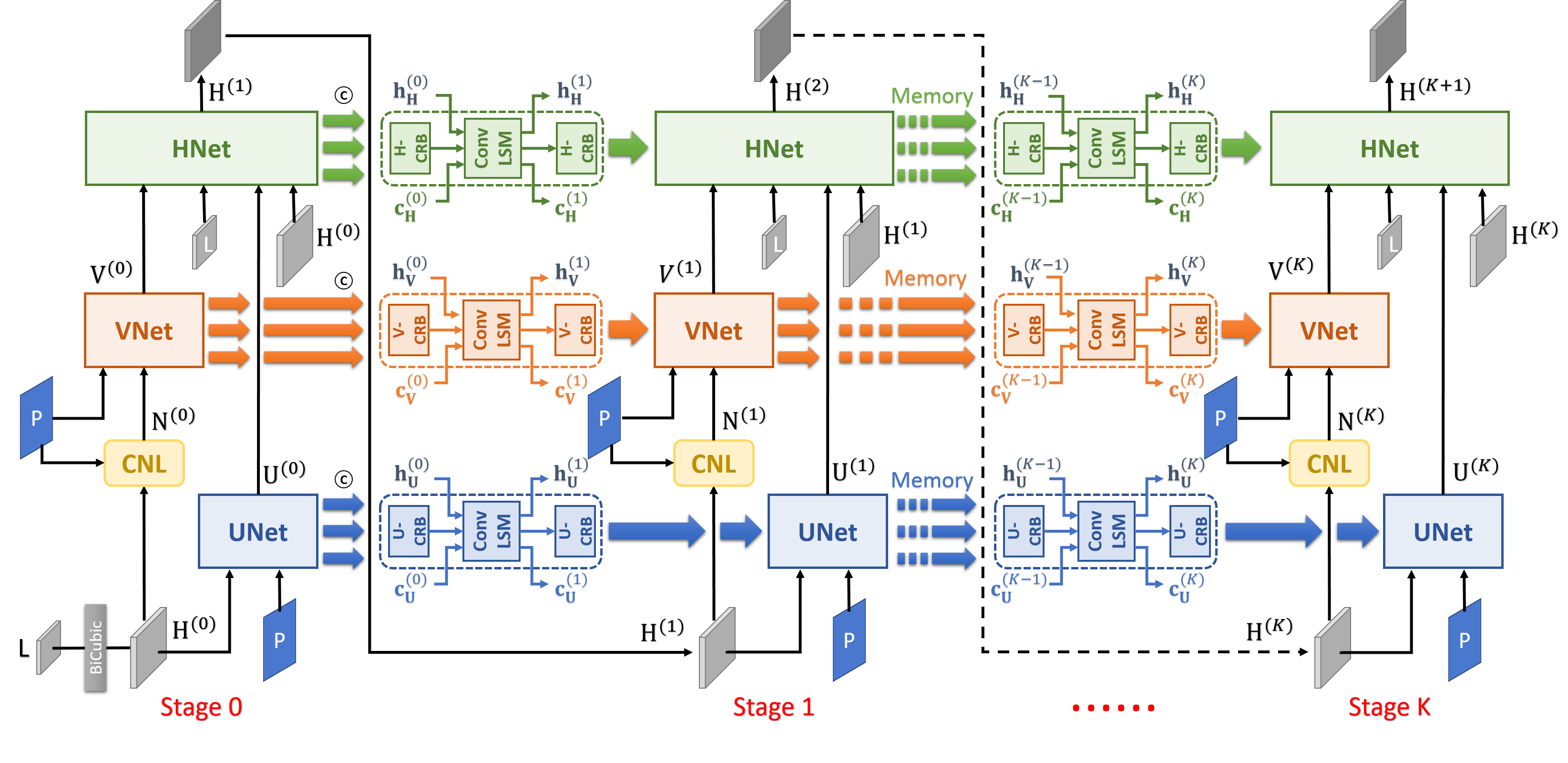}
\caption{The overall architecture of our proposed unfolding network, consisting of information flow and memory flow. For information flow, the LR target image $\mathbf{L}$ is firstly up-sampled as $\mathbf{H}^{(0)}$ and then performs the stage-wise iteration updating in the overall ${\rm K}$ stages where ${\rm UNet}$ and ${\rm VNet}$ are parallel updated and then transmitted into ${\rm HNet}$. To emphasize, ${\rm CNL}$ module aims to explore the cross-modality information to transfer the detailed structures of guidance image into the target image to generate the immediate output $\mathbf{N}$. To facilitate the signal flow across iterative stages, the persistent memory mechanism (remarked as memory flow) is introduced to augment the information representation by exploiting  the  Long  short-term  memory  unit  (LSTM)  in  the image and feature spaces. }
\label{mainfig}
\end{center}
\end{figure*}

\vspace{-.2cm}
\section{Proposed Approach}\label{sec:proposed}
\vspace{-.2cm}
In this section, we provide a detailed introduction to our proposed memory-augmented deep unfolding GISR network, illustrated in Fig.~\ref{mainfig}. For convenience, we first define some notations used in the GISR model. To be specific, $\mathbf{L}\in\mathbb{R}^{m\times n \times B}$ denotes the low-resolution (LR) target image, $\mathbf{H} \in \mathbb{R}^{M\times N \times B}$ represents the corresponding high-resolution (HR) target image, and $\mathbf{P} \in \mathbb{R}^{M\times N \times b}$ is the guidance image. 

\vspace{-.2cm}
\subsection{MAP Model for Guided Image Super-resolution}
In GISR, we assume that the LR target image $\mathbf{L}$ is obtained through performing the blurring kernel $\mathbf{k}$ and down-sampling operator over the HR target image $\mathbf{H}$, and thus the degradation model can be mathematically formulated as
\begin{equation}\label{eq:ab}
     \mathbf{L} = (\mathbf{H}\ast \mathbf{k})\downarrow_{\mathbf{s}} + \mathbf{n_s},  
\end{equation}
where $\ast$ represents the convolution operation, and $\mathbf{n_s}$ is usually assumed to be additive white Gaussian noise (AWGN) with standard deviation $\sigma$. The spatial resolution ratio between $\mathbf{H}$ and $\mathbf{L}$ is $r=M/m=N/n$. The observation model in Eq.~(\ref{eq:ab}) can be equivalently reformulated as
\begin{equation}\label{eq:abmatrix}
     \mathbf{L} = \mathbf{D}\mathbf{K}\mathbf{H}+ \mathbf{n_s},  
\end{equation}
where $\mathbf{K}$ is the matrix form of kernel $\mathbf{k}$, and $\mathbf{D}$ is the matrix form of down-sampling operator. Based on the observation model in Eq.~(\ref{eq:abmatrix}), the distribution of $\mathbf{L}$ is defined as 
\begin{equation}\label{eq:likelihood}
     P(\mathbf{L}|\mathbf{H})=\mathcal{N}(\mathbf{L}|\mathbf{D}\mathbf{K}\mathbf{H},\sigma^{2}\mathbf{I}),
\end{equation}
where $\mathcal{N}(\mathbf{L}|\mathbf{D}\mathbf{K}\mathbf{H},\sigma^{2}\mathbf{I})$ denotes the Gaussian distribution with mean $\mathbf{D}\mathbf{K}\mathbf{H}$ and covariance matrix $\sigma^{2}\mathbf{I}$.

\begin{figure}
\begin{center}
\includegraphics[width=0.35\textwidth]{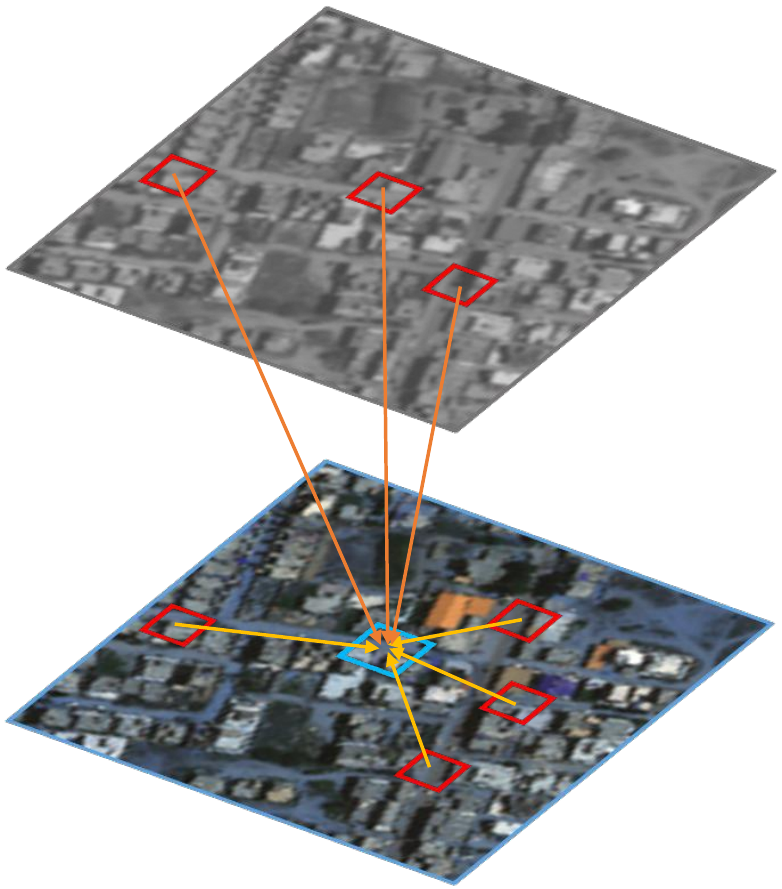}
\caption{Illustration of non-local cross-modality aggregation. Since the target and guidance image capture the same scene, consistent patterns thus exist between the two images. The HR target image can be reconstructed by aggregating the long-range correlation information from both target and guidance images.}
\label{aggregation}
\end{center}
\end{figure}

Since GISR super-resolves the LR target image under the guidance image, which is usually captured in the same scene with the LR target image, the LR target image and the guidance image thus share some global and local relevant features. Existing GISR methods only consider one of the two shared features. To capture both features, we design two types of image priors, i.e., local implicit prior and global implicit prior. The local implicit prior models the relationship between the HR target image and the HR guidance image implicitly from a local perspective, and thus can help capture the local related information between the target image and the guidance image. The global implicit prior considers the non-local auto-regression property between the two images from a global perspective, and thus the global correlation between the two images can be well exploited. Since the target and guidance image capture the same scene, consistent patterns thus exist between the two images. The HR target image can be reconstructed by aggregating the long-range correlation information from both target and guidance images as shown in Fig.~\ref{aggregation}.

Specifically, we assume the local implicit prior distribution $P_{1}(\mathbf{H}|\mathbf{P})$ and global implicit prior distribution $P_{2}(\mathbf{H}|\mathbf{P})$ separately as follows: 
\begin{eqnarray}\label{eq:priors}
     P_{1}(\mathbf{H}|\mathbf{P})&\propto&\exp\{-\eta\Omega_1(\mathbf{H}|\mathbf{P})\},\\
     P_{2}(\mathbf{H}|\mathbf{P})&\propto&\exp\{-\lambda\Omega_{NL}(\mathbf{H}|\mathbf{P})\},
\end{eqnarray}
where $\Omega_1(\mathbf{H}|\mathbf{P})$ and $\Omega_{NL}(\mathbf{H}|\mathbf{P})$ are two energy functions related to $\mathbf{H}$ and $\mathbf{P}$, $\eta$ and $\lambda$ are the weight parameters. For simplicity, we assume the distribution of $\mathbf{H}$ is
\begin{eqnarray}\label{eq:Hpriors}
     P(\mathbf{H}|\mathbf{P})\propto P_{1}(\mathbf{H}|\mathbf{P})P_{2}(\mathbf{H}|\mathbf{P}),
\end{eqnarray}
Therefore, the posterior of $\mathbf{H}$ given $\mathbf{L}$ and $\mathbf{P}$ can be computed by the Bayes formula:
\begin{eqnarray}\label{eq:posterior}
     P(\mathbf{H}|\mathbf{L},\mathbf{P})=\frac{P(\mathbf{L}|\mathbf{H})P(\mathbf{H}|\mathbf{P})}{P(\mathbf{L}|\mathbf{P})},
\end{eqnarray}
where $P(\mathbf{L}|\mathbf{P})$ is the marginal distribution of $\mathbf{L}$ which is not related with $\mathbf{H}$. By using the maximum a posterior (MAP) principle, $\mathbf{H}$ can be obtained by maximizing the log-posterior $\log P(\mathbf{H}|\mathbf{L},\mathbf{P})$, which is equivalent to the following optimization problem:
\begin{eqnarray}\label{eq:MAP}
     \max_{\mathbf{H}}\log P(\mathbf{L}|\mathbf{H})+\log P_{1}(\mathbf{H}|\mathbf{P}) + \log P_{2}(\mathbf{H}|\mathbf{P}).
\end{eqnarray}
Further, Eq.~(\ref{eq:MAP}) can be reformulated as 
\begin{eqnarray}\label{eq:model}
     \max\limits_{\mathbf{H}}\frac{1}{2}\left|\left|\mathbf{L}-\mathbf{D}\mathbf{K}\mathbf{H}\right|\right|_2^2+\eta\Omega_1(\mathbf{H}|\mathbf{P}) + \lambda\Omega_{NL}(\mathbf{H}|\mathbf{P}).
\end{eqnarray}
Eq.~(\ref{eq:model}) is our final proposed model. In the next section, we will develop an optimization algorithm to solve this model. 

\subsection{Model Optimization}\label{sec:3architecture}
We now solve the optimization problem of Eq.~(\ref{eq:model}) using half-quadratic splitting (HQS) algorithm, which has been widely used in solving image inverse problems~\citep{geman1992constrained,geman1995nonlinear,krishnan2009fast,6518114}. By introducing two auxiliary variables $\mathbf{U}$ and $\mathbf{V}$, Eq.~(\ref{eq:model}) can be reformulated as a non-constrained optimization problem:
\begin{align}
\mathop{\min}\limits_{\mathbf{H},\mathbf{U},\mathbf{V}} 
    &\frac{1}{2}\left|\left|\mathbf{L}-\mathbf{D}\mathbf{K}\mathbf{H}\right|\right|_2^2+
    \frac{\eta_1}{2} \left|\left|\mathbf{U}-\mathbf{H}\right|\right|_2^2+\eta\Omega_1(\mathbf{U}|\mathbf{P}) 
    \notag \\
    & +\frac{\lambda_1}{2}\left|\left|\mathbf{V}-\mathbf{H}\right|\right|_2^2+\lambda\Omega_{NL}(\mathbf{V}|\mathbf{P}) \label{eq:min_opt3}
\end{align}
where $\eta_1$ and $\lambda_1$ are penalty parameters. When $\eta_1$ and $\lambda_1$ simultaneously approach to infinity, the solution of minimizing Eq.~(\ref{eq:min_opt3}) converges to that of minimizing Eq.~(\ref{eq:model}). Then, minimizing Eq.~(\ref{eq:min_opt3}) can be achieved
by solving three sub-problems for alternately updating $\mathbf{U}$, $\mathbf{V}$, and $\mathbf{H}$.

\textbf{Updating $\mathbf{U}$.} Given the estimated HR target image $\mathbf{H}^{(k)}$ at iteration $k$, the auxiliary variable $\mathbf{U}$ can be updated as:
\begin{align}
    \mathbf{U}^{(k)}  = \mathop{\arg\min}\limits_{\mathbf{U}} \frac{\eta_1}{2}\left|\left|\mathbf{U}-\mathbf{H}^{(k)}\right|\right|_2^2+\eta_2\Omega_1(\mathbf{U}|\mathbf{P}).   \label{eq:opt_u11}
\end{align}
By applying the proximal gradient method~\citep{1976Monotone} to Eq.~(\ref{eq:opt_u11}), we can derive
\begin{align}
    \mathbf{U}^{(k)}=  {\rm prox}_{\Omega_1(.)}(\mathbf{U}^{(k-1)}-\delta_1\nabla f_1(\mathbf{U}^{(k-1)}))      \label{uk1}
\end{align}
where ${\rm prox}_{\Omega_1}(\cdot)$ is the proximal operator corresponding to the implicit prior $\Omega_1(\cdot)$, $\delta_1$ denotes the updating step size, and the gradient $\nabla f_1(\mathbf{U}^{(k-1)})$ is
\begin{align}
    \nabla f_1(\mathbf{U}^{(k-1)})= \mathbf{U}^{(k-1)}-\mathbf{H}^{(k)}. 
\end{align}

\textbf{Updating $\mathbf{V}$.} Given $\mathbf{H}^{(k)}$, $\mathbf{V}$ can be updated as:
\begin{align}
       \mathbf{V}^{(k)}  = \mathop{\arg\min}\limits_{\mathbf{V}} \frac{\lambda_1}{2}\left|\left|\mathbf{V}-\mathbf{H}^{(k)}\right|\right|_2^2+\lambda\Omega_{NL}(\mathbf{V}|\mathbf{P}).    \label{eq:opt_v11} 
\end{align}
Similarly, we can obtain
\begin{align}
    \mathbf{V}^{(k)}=  {\rm prox}_{\Omega_{NL}(.)}(\mathbf{V}^{(k-1)}-\delta_2\nabla f_2(\mathbf{V}^{(k-1)}))    \label{vk}
\end{align}
where ${\rm prox}_{\Omega_{NL}}(\cdot)$ is the proximal operator corresponding to the non-local prior term $\Omega_{NL}(\cdot)$, $\delta_2$ indicates the updating step size, and the gradient $\nabla f_2(\mathbf{V}^{(k-1)})$ is computed as
\begin{align}
    \nabla f_2(\mathbf{V}^{(k-1)})= \mathbf{V}^{(k-1)}-\mathbf{H}^{(k)}.
    \label{grad_v}
\end{align}

\textbf{Updating $\mathbf{H}$.} Given $\mathbf{U}^{(k)}$ and $\mathbf{V}^{(k)}$, $\mathbf{H}$ is updated as:
\begin{align}
    \mathbf{H}^{(k+1)}  &=\mathop{\arg\min}\limits_{\mathbf{H}} 
    \frac{1}{2}\left|\left|\mathbf{L}-\mathbf{D}\mathbf{K}\mathbf{H}\right|\right|_2^2 
      +\frac{\eta_1}{2}\left|\left|\mathbf{U}^{(k)}-\mathbf{H}\right|\right|_2^2 \notag \\ 
      & +\frac{\lambda_1}{2}\left|\left|\mathbf{V}^{(k)}-\mathbf{H}\right|\right|_2^2.  \label{eq:opt_h1}
\end{align}
Although we can derive closed form solution of updating $\mathbf{H}$ from Eq.~(\ref{eq:opt_h1}), the updating equation relies on calculating the inverse of a large matrix, which is computational inefficiency. To alleviate this issue, we still follow the updating rules of $\mathbf{U}$ and $\mathbf{V}$ and adopt the gradient decent method to update $\mathbf{H}$. Therefore, the updating equation for $\mathbf{H}$ is
\begin{align}
    \mathbf{H}^{(k+1)}= \mathbf{H}^{(k)}-\delta_3\nabla f_3(\mathbf{H}^{(k)}), \label{hk1}
\end{align}
where $\delta_3$ is the step size, and the gradient $\nabla f_3(\mathbf{H}^{(k)})$ is 
\begin{align}
    \nabla f_3(\mathbf{H}^{(k)}) &= (\mathbf{D}\mathbf{K})^T(\mathbf{D}\mathbf{K}\mathbf{H}^{(k)}-\mathbf{L})  + \eta_1(\mathbf{H}^{(k)}-\mathbf{U}^{(k)}) \notag \\
    & + \lambda_1(\mathbf{H}^{(k)}-\mathbf{V}^{(k)})   \label{eq:iter_gh2}
\end{align}
where $T$ is the matrix transpose operation. 

\begin{figure*}
\begin{center}
\includegraphics[width=\textwidth]{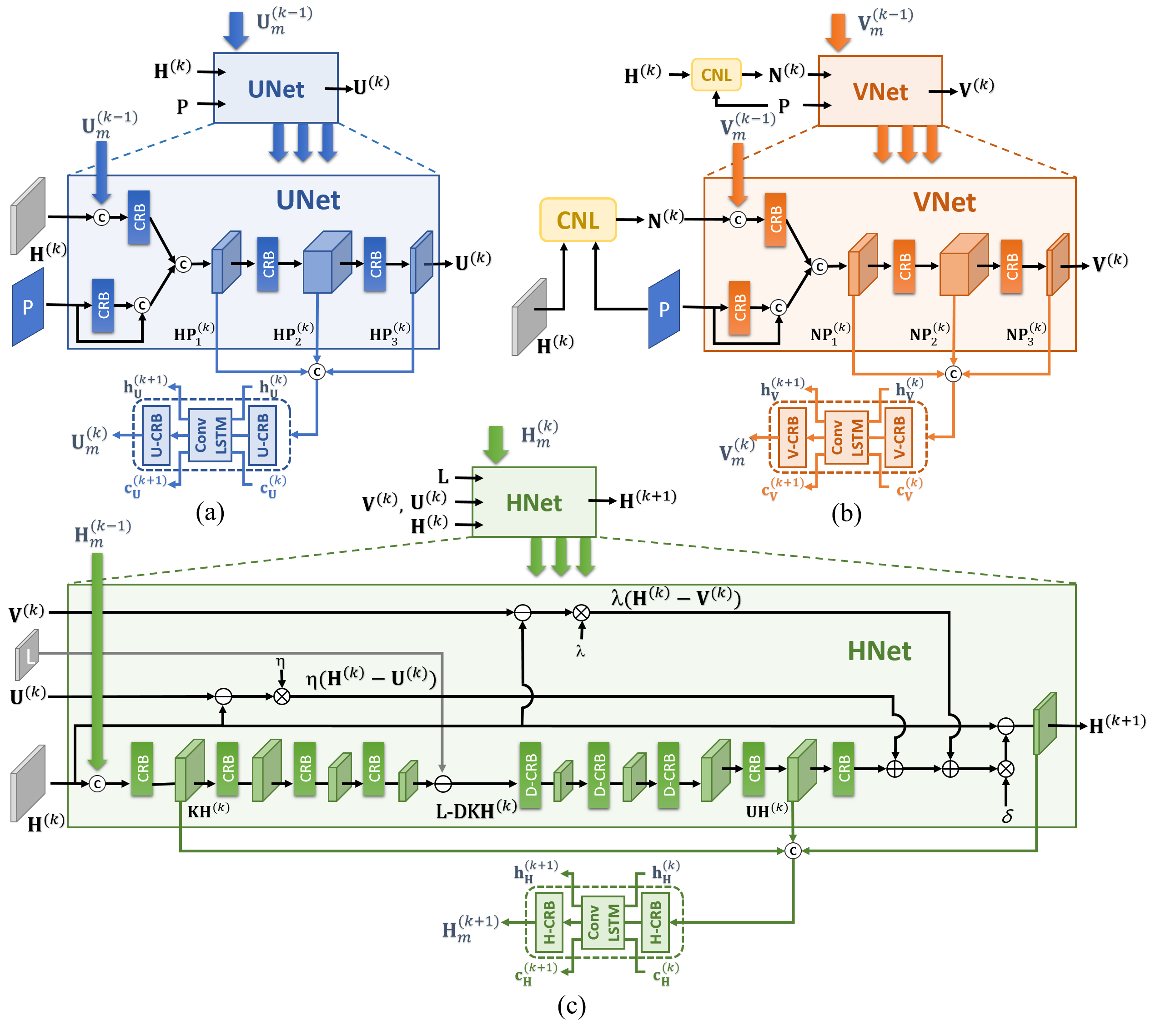}
\caption{The detailed architecture of ${\rm UNet}$, ${\rm VNet}$, and ${\rm HNet}$ in stage $k$ of the overall network. Each module corresponds to one operation of the iterative step for updating $\mathbf{U}$, $\mathbf{V}$, and $\mathbf{H}$.}
\label{sonfig}
\end{center}
\end{figure*}

\subsection{Deep unfolding network} \label{DUN}
Based on the iterative algorithm, we build a deep neural network for GISR as illustrated in
Fig.~\ref{mainfig}. This network is an implementation of the algorithm for solving Eq.~(\ref{eq:model}). Each stage receives the inputs $\mathbf{U}^{(k-1)}$, $\mathbf{V}^{(k-1)}$, and $\mathbf{H}^{(k)}$, and generates the outputs $\mathbf{U}^{(k)}$, $\mathbf{V}^{(k)}$, and $\mathbf{H}^{(k+1)}$.

In the proposed algorithm, the two proximal operators ${\rm prox}_{\Omega_1}(\cdot)$ and ${\rm prox}_{\Omega_{NL}}(\cdot)$ can not be explicitly deduced since the regularization terms $\Omega_1(\cdot)$ and $\Omega_{NL}(\cdot)$ are not explicitly defined. We thus use deep CNNs to learn the two proximal operators for updating $\mathbf{U}^{(k-1)}$ and $\mathbf{V}^{(k-1)}$. 

Specifically, we can easily learn ${\rm prox}_{\Omega_1}(\cdot)$ in Eq.~(\ref{uk1}) using the CNN, dubbed as ${\rm UNet}$ as follows:
\begin{eqnarray}
    \mathbf{U}^{(k)} &= {\rm prox}_{\Omega_1}(\mathbf{U}^{(k-1)}, \mathbf{H}^{(k)})\nonumber\\
     &\approx{\rm UNet}(\mathbf{U}^{(k-1)}, \mathbf{H}^{(k)}, \mathbf{P}),\label{uprox}
\end{eqnarray}
where $\mathbf{P}$ is added as an input of the ${\rm UNet}$ due to the regularization term $\Omega_1(\mathbf{U}|\mathbf{P})$ in Eq.~\ref{eq:opt_u11} that implies $\mathbf{U}$ and $\mathbf{P}$ have some relations. ${\rm UNet}$ represents the CNN whose input and output have the same spatial resolution. The detailed implementation of ${\rm UNet}$ is illustrated in Fig.~\ref{sonfig} (a). 

\begin{figure}
\begin{center}
\includegraphics[width=0.45\textwidth]{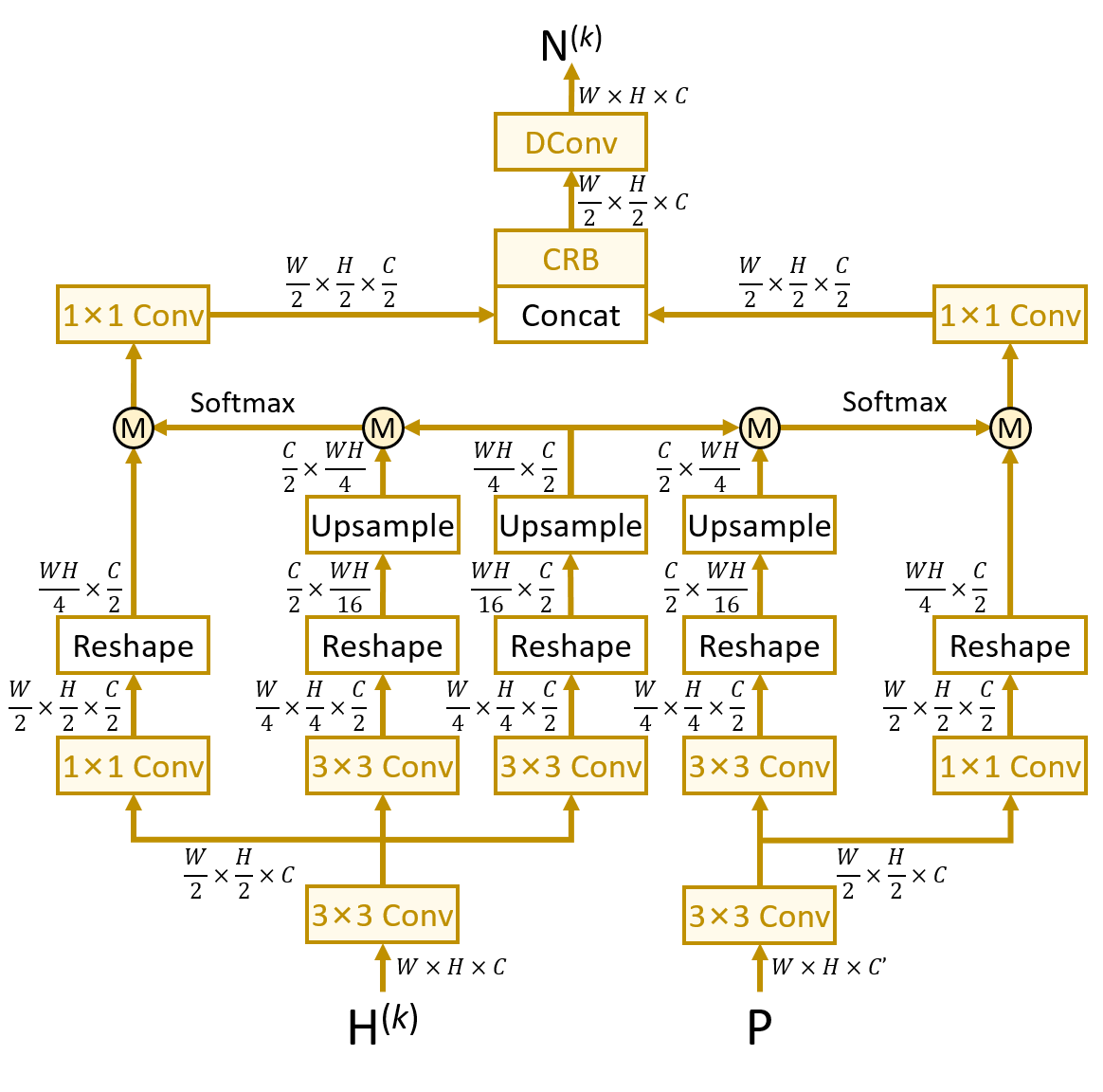}
\caption{The cross-modality non-local operation module. It takes the updated HR target image $\mathbf{H}^{(k)}$ and guidance image $\mathbf{P}$ as input and generates the refined image $\mathbf{N}^{(k)}$, referred to Eq.~(\ref{ukn}).}
\label{non-local}
\end{center}
\end{figure}

Before learning ${\rm prox}_{\Omega_{NL}}(\cdot)$ using the CNN, we firstly introduce an intermediate variable $\mathbf{N}^{(k)}$ to explicitly consider the non-local term $\Omega_{NL}(\mathbf{H}|\mathbf{P})$ by devising a cross-modalities non-local module, denoted as ${\rm CNL}$, which takes the $\mathbf{H}^{(k)}$ and $\mathbf{P}$ as input and then generates the output value $\mathbf{N}^{(k)}$. The ${\rm CNL}$ module is defined as 
\begin{align}
    \mathbf{N}^{(k)}=  {\rm CNL}(\mathbf{H}^{(k)},\mathbf{P}),      \label{ukn}
\end{align}
and is illustrated in Fig.~\ref{non-local}. The output $\mathbf{N}^{(k)}$ of ${\rm CNL}$ module can be interpreted as a slight fine-tuning on the HR target image $\mathbf{H}^{(k)}$ using the guided image $\mathbf{P}$. Then, $\mathbf{N}^{(k)}$ is plugged into Eq.~(\ref{grad_v}) to replace $\mathbf{H}^{(k)}$ for computing the gradient $\nabla f_2(\mathbf{V}^{(k-1)})$. Finally, we can also learn ${\rm prox}_{\Omega_{NL}}(\cdot)$ in Eq.~(\ref{vk}) using the CNN, dubbed as ${\rm VNet}$ as follows:
\begin{eqnarray}
    \mathbf{V}^{(k)} &= {\rm prox}_{\Omega_{NL}}(\mathbf{V}^{(k-1)}, \mathbf{H}^{(k)})\nonumber\\
     &\approx{\rm VNet}(\mathbf{V}^{(k-1)}, \mathbf{N}^{(k)}, \mathbf{P}),
\end{eqnarray}
where ${\rm VNet}$ has similar structure with ${\rm UNet}$ and its implementation is illustrated in Fig.~\ref{sonfig} (b). 

To implement Eq.~(\ref{hk1}) and Eq.~(\ref{eq:iter_gh2}) using the network, we firstly split Eq.~(\ref{eq:iter_gh2}) into three steps:
\begin{align}
    \hat{\mathbf{L}}^{(k)} = & \mathbf{D}\mathbf{K}\mathbf{H}^{(k)}             \label{s11}\\
    \mathbf{E}^{(k)} = & (\mathbf{D}\mathbf{K})^T(\hat{\mathbf{L}}^{(k)}-\mathbf{L})      \label{s12}\\  
    \mathbf{R}^{(k)} = & \mathbf{E}^{(k)} + \eta_1(\mathbf{H}^{(k)}-\mathbf{U}^{(k)}) + \lambda_1(\mathbf{H}^{(k)}-\mathbf{V}^{(k)}).             \label{s13}
\end{align}
These steps can be transformed into a network, dubbed as ${\rm HNet}$, containing many modules corresponding to each operation in the three steps. To be specific, given the $k$-iteration approximated HR target image $\mathbf{H}^{(k)}$, Eq.~(\ref{s11}) generates an immediate LR version $\hat{\mathbf{L}}^{(k)}$ by implementing the down-sampling $\mathbf{D}$ and low-passing filtering $\mathbf{K}$ functions. This module is defined as 
\begin{align}
    \hat{\mathbf{L}}^{(k)} = Down\downarrow_s(\mathbf{H}^{(k)})             \label{si1}
\end{align}
where $Down\downarrow_s$ denotes the CNN with the spatial resolution reduction by $s$ times and the reduction is conducted by convolution operator with the $s$ strides. 

Followed by, Eq.~(\ref{s12}) first computes the LR residual between the input LR target image $\mathbf{L}$ and the generated one $\hat{\mathbf{L}}^{(k)}$, and then acquires the HR residuals $\mathbf{E}^{(k)}$ by applying the corresponding transpose operation $(\mathbf{D}\mathbf{K})^T$ to the LR residual. In detail, the transpose function is implemented by the deconvolution filters with $s$ strides as follows:
\begin{align}
    \hat{\mathbf{E}}^{(k)} = Up\uparrow_s(\hat{\mathbf{L}}^{(k)},\mathbf{L}).             \label{si2}
\end{align}
Finally, by combining the Eq.~(\ref{s13}) and Eq.~(\ref{hk1}), it is trivial to generate the $k+1$-iteration HR target image $\mathbf{H}^{(k+1)}$ in context of $\mathbf{U}^{(k)}$ and $\mathbf{V}^{(k)}$ as follows:
\begin{align}
    \mathbf{H}^{(k+1)}= \mathbf{H}^{(k)}-\delta_3 \mathbf{R}^{(k)}.    \label{eq:iter_h}
\end{align}
To this end, the obtained $\mathbf{U}^{(k)}$, $\mathbf{V}^{(k)}$, and $\mathbf{H}^{(k+1)}$ are transmitted into the next stage. 

In summary, the proposed algorithm can be approximated by the unfolding network shown in Fig.~\ref{mainfig}. The signal flow of the deep unfolding network is
\begin{align}
    &\mathbf{H}^{(k)}\Rightarrow \mathbf{U}^{(k)}, \notag \\
    &\mathbf{H}^{(k)}\Rightarrow \mathbf{N}^{(k)} \Rightarrow \mathbf{V}^{(k)}, \notag \\
    &[\mathbf{U}^{(k)}, \mathbf{V}^{(k)}] \Rightarrow \mathbf{H}^{(k+1)}. \label{oppipe}
\end{align}
In each stage, the network achieves the sequential updates of the auxiliary variable $\mathbf{U}$, the auxiliary variable $\mathbf{N}$, the auxiliary variable $\mathbf{V}$ and the approximated HR target $\mathbf{H}$, as illustrated in Eq.~\ref{oppipe}. In each stage, since each module corresponds to the operation in one iteration step, the interpretability of the deep model can thus be improved. 

\subsection{Memory-Augmented Deep Unfolding Network}
Nevertheless, there exist several issues of deep unfolding network to be solved. First, the potential of cross-stages, which can be regarded as short-term memory has not been fully explored. In addition, the severe information loss between adjacent stages, recognized as the rarely realized long-term dependency has not been studied due to the feature transformation with channel number reduction, further limiting their improvements. 
 
In this paper, to facilitate the signal flow across iterative stages, the persistent memory mechanism is introduced to augment the information representation by exploiting the long-short term unit in the image and feature spaces. Specifically, based on the proposed unfolding network in the previous section, we further embed the memory mechanism into it. In the three sub-networks (i.e., ${\rm UNet}$, ${\rm VNet}$, ${\rm HNet}$), the different-layers feature maps  and  the  output  intermediate  images of each iterative module are selected and integrated for further transformation and then inserted into the next iterative stage for information interaction across stages, thus reducing the information loss. In this way, the information representation can be well improved, leading to better performance. Equipped with above persistent memory mechanism, we propose a memory-augmented deep unfolding network (MADUNet) for GISR as shown in Fig.~\ref{mainfig}. Next, we will introduce the improved version of ${\rm UNet}$, ${\rm VNet}$, and ${\rm HNet}$ with ebedded memory mechanism in detail.

\subsubsection{{\rm UNet}} 
To increase the model capability, the memory of previous information at previous stages is introduced to the expressed module corresponding to Eq.~(\ref{uprox}). As shown in Fig.~\ref{sonfig}, the ${\rm UNet}$ is designed with the basic ${\rm CRB}$ unit  which consists of the pure convolution layer and the effective residual blocks. Taking the $k$-th iteration for example, the computation flow of ${\rm UNet}$ is defined as
\begin{align}
\centering
    &\mathbf{P}_1^{(k-1)} = {\rm Cat}({\rm CRB}(\mathbf{P}),\mathbf{P}) \\ 
    &\mathbf{H}_1^{(k-1)}  = {\rm Cat}({\rm CRB}(\mathbf{H}^{(k-1)}),\mathbf{U}_m^{(k-1)})\\
    &\mathbf{HP}_1^{(k-1)} = {\rm Cat}(\mathbf{H}_1^{(k-1)},\mathbf{P}_1^{(k-1)})\\
    &\mathbf{HP}_2^{(k-1)}  = {\rm CRB}(\mathbf{HP}_1^{(k-1)})\\
    &\mathbf{U}^{(k)}  = {\rm CRB}(\mathbf{HP}_2^{(k-1)}),
\end{align}  
where ${\rm Cat}$ represents the concatenation operation along the channel dimension and $\mathbf{U}_m^{(k-1)}$ is the high-throughput information from previous stage to reduce the information loss. The updated memory $\mathbf{U}^{(k)}$ can be obtained by exploiting ${\rm ConvLSTM}$ unit to transform the different-layer's features $\mathbf{HP}_1^{(k-1)}$, $\mathbf{HP}_2^{(k-1)}$  and $\mathbf{U}^{(k)}$ as
\begin{align}
\centering
    &\mathbf{HPU} = {\rm CRB}({\rm Cat}(\mathbf{HP}_1^{(k-1)},\mathbf{HP}_2^{(k-1)},\mathbf{U}^{(k)})) \\
    &\mathbf{h}_\mathbf{U}^{(k)}, \mathbf{c}_\mathbf{U}^{(k)} = {\rm ConvLSTM}(\mathbf{HPU},\mathbf{h}_\mathbf{U}^{(k-1)},\mathbf{c}_\mathbf{U}^{(k-1)}) \\
    & \mathbf{U}_m^{(k)} = {\rm CRB}(\mathbf{h}_\mathbf{U}^{(k)})
\end{align}  
where $\mathbf{h}_\mathbf{U}^{(k-1)}$ and $\mathbf{c}_\mathbf{U}^{(k-1)}$ denotes the hidden state and cell state in $ConvLSTM$ to augment the long-range cross stage information dependency. Furthermore, $\mathbf{h}_\mathbf{U}^{(k)}$ is directly fed into the ${\rm CRB}$ to generate the updated memory $\mathbf{U}_m^{(k)}$. The transition process of ${\rm ConvLSTM}$ is unfolded as
\begin{align}
\centering
&\mathbf{i}^{(k)}= \sigma(\mathbf{W}_{si}\ast \mathbf{HPU}+ \mathbf{W}_{hi}\ast \mathbf{h}_\mathbf{U}^{(k-1)} +\mathbf{b}_i), \\
&\mathbf{f}^{(k)} = \sigma(\mathbf{W}_{sf}\ast \mathbf{HPU}+ \mathbf{W}_{hf}\ast \mathbf{h}_\mathbf{U}^{(k-1)} +\mathbf{b}_f), \\
& \mathbf{c}^{(k)} = \mathbf{f}^{(k)}\odot \mathbf{c}_\mathbf{U}^{(k-1)} + \mathbf{i}^{(k)} \odot tanh(\mathbf{W}_{sc}\ast \mathbf{HPU}\\ \nonumber
&+ \mathbf{W}_{hc}\ast \mathbf{h}_\mathbf{U}^{(k-1)} +\mathbf{b}_c), \\
& \mathbf{o}^{(k)} = \sigma(\mathbf{W}_{so}\ast \mathbf{HPU}+ \mathbf{W}_{ho}\ast \mathbf{h}_\mathbf{U}^{(k-1)} + \mathbf{b}_o), \\
 & \mathbf{h}_\mathbf{U}^{(k)} = \mathbf{o}^{(k)} \odot tanh(\mathbf{c}_\mathbf{U}^{(k)})
\end{align}
where $\ast$ and $\odot$ denote the convolution operation and Hadamard product, respectively. $\mathbf{c}_\mathbf{U}^{(k)}$ and $\mathbf{h}_\mathbf{U}^{(k)}$ represent the cell state and hidden state, respectively. $\sigma$ and $tanh$ denote the sigmoid and tanh function, respectively. In this way, not only the information loss of feature channel reduction is alleviated, but also the long-term cross-stage information dependency can be enhanced.  

\subsubsection{{\rm VNet}} 
In Eq.~\ref{ukn}, it aims to measure the non-local cross-modality similarity and then aggregates the semantically-related and structure-consistent content from long-range patches in target images and across the modalities of target and guidance images.   To this end, we devise a novel cross-modality non-local operation module (denoted as ${\rm CNL}$). Fig.~\ref{non-local} illustrates the ${\rm CNL}$ module, which receives the updated HR target image $\mathbf{H}^{(k)}$ and guidance image $\mathbf{P}$ as input and generates the refined image $\mathbf{N}^{(k)}$.
	
Specifically, the updated HR target image feature map $\mathbf{H}^{(k)}$ with the size of $W \times H \times C$ and guidance image feature map $\mathbf{P}$ with the size of $W \times H \times C$ are transmitted into the ${\rm CNL}$. Firstly, we employ two independent $3 \times 3$ convolution over the target and guidance features, thus reducing the input feature dimension of $\mathbf{H}^{(k)}$ and $\mathbf{P}$ to $\frac{W}{2} \times \frac{H}{2} \times C$, which is the dimension of $\mathbf{H}_r$ and $\mathbf{P}_r$. Secondly, to model the long-range correlation, both the cross-modality and inter-modality operations are conducted. In the left part of Fig.~\ref{non-local}, the inter-modality computing imitates the non-local mean filtering over target image feature maps as follows:
	\begin{align}
	&\mathbf{H}_{r1}=\delta(\mathbf{H}_r),\\
	&\mathbf{H}_{r2}=\theta(\mathbf{H}_r),\\ \label{hn2}
	&\mathbf{F}_{HH} = {\rm softmax}(\mathbf{H}_{r1} \mathbf{H}_{r2}),
	\end{align}
where the target features $\mathbf{H}_r$ is further processed by the $\delta$ and $\theta$ convolution modules separately, which contains the sequential operation of $3\times 3$ convolution, the reshape and the near interpolation up-sampling operations, thus generating the two features $\mathbf{H}_{r1}$ and $\mathbf{H}_{r2}$ with size of $\frac{C}{2} \times \frac{WH}{4}$ and $\frac{WH}{4} \times \frac{C}{2}$. Then, the multiplication of $\mathbf{H}_{r1}$ and $\mathbf{H}_{r2}$ are input to a {\rm softmax} function to generate the attention map $\mathbf{F}_{HH}$. 

To exploit the cross-modality correlation, guidance image feature $\mathbf{P}_r$ is passed into the $\phi$ convolution module to obtain the feature $\mathbf{P}_{r1}$ as follows:
	\begin{equation}
	\mathbf{P}_{r1} = \phi(\mathbf{P}_r), \label{pn1}
	\end{equation}
Then, the cross-modality correlation can be modeled as
	\begin{align}
	&\mathbf{F}_{HP} = {\rm softmax}(\mathbf{H}_{r2} \mathbf{P}_{r1}).
	\end{align}
Finally, the additional $\varphi(.)$ and $\vartheta(.)$ modules are adopted over $\mathbf{H}^{(k)}$ and $\mathbf{P}$ to provide the the embedding representation $\mathbf{H}_{e}$ and $\mathbf{P}_e$ with the same size $\frac{WH}{4} \times \frac{C}{2}$. Incorporating the inter-modality $\mathbf{F}_{HH}$ and cross-modality $\mathbf{F}_{HP}$ correlation, the refined feature map $\mathbf{N}^{(k)}$ can be formulated as
\begin{align}
	&\mathbf{N}^{(k)}= {\rm CRB}({\rm Cat}({\rm C_1}(\mathbf{F}_{HP}\mathbf{P}_e), {\rm C_1}(\mathbf{F}_{HH}\mathbf{H}_e))),
\end{align}
where ${\rm C_1}(.)$ is the the convolution layer with $1 \times 1$ kernel size. With the proposed ${\rm CNL}$, it is capable of searching the similarities between long-range patches in target  images and across the modalities of target and guidance images, benefiting the texture enhancement.

With the output of CNL module $\mathbf{N}^k$, the previous output $\mathbf{V}^{(k-1)}$ and the accumulated memory state $\mathbf{V}_m^{(k-1)}$, we can obtain the updated $\mathbf{V}^{(k)}$ as shown in Fig.~\ref{sonfig} (b). It can be clearly seen that the ${\rm VNet}$ has a similar architecture with that of ${\rm UNet}$, which is consistent with their similar updating rules. Additionally, the memory transmission of ${\rm VNet}$ is also the same as that of ${\rm UNet}$.

\subsubsection{{\rm HNet}} 
To transform the update process of $\mathbf{H}^{(k+1)}$, i.e., Eq.~(\ref{s13}), Eq~(\ref{si1}), and Eq.~(\ref{si2}), into a network. Firstly, we need to implement the two operations, i.e.,  $Down\downarrow_s$ and $Up\uparrow_s$, using the network. Specifically, $Down\downarrow_s$ is implemented by a ${\rm CRB}$ module with spatial identify transformation, and an additional $s$-strides followed ${\rm CRB}$ module with spatial resolution reduction: 
\begin{align}
\centering
    &\mathbf{KH}^{(k)} = {\rm CRB}({\rm Cat}(\mathbf{H}^{(k)},\mathbf{H}_m^{(k)})) \\
    &\mathbf{DKH}^{(k)} = {\rm CRB}^{(s)}\downarrow(\mathbf{KH}^{(k)}) \label{hdown}
\end{align}
where ${\rm CRB}^{(s)}\downarrow$ aims to perform the $s$ times down-sampling. The latter operation $Up\uparrow_s$ is implemented by a deconvolution layer containing the $s$-strides ${\rm CRB}$ module with spatial resolution expansion and a ${\rm CRB}$ module with spatial identify transformation:
\begin{align}
\centering
    &\mathbf{UH}^{(k)} = {\rm CRB}^{(s)}\uparrow(\mathbf{L}-\mathbf{DKH}^{(k)})  \label{hup}
\end{align}
where ${\rm CRB}^{(s)}\uparrow$ aims to perform the $s$ times up-sampling. Further, in context of Eq.~(\ref{eq:iter_h}), Eq.~(\ref{hdown}) and Eq.~(\ref{hup}), the updated $\mathbf{H}^{(k+1)}$ and the updated memory $\mathbf{H}_m^{(k+1)}$ can be obtained as follows:
\begin{align}
\centering
    &\mathbf{MH}^{(k+1)} = {\rm CRB}({\rm Cat}(\mathbf{KH}^{(k)},\mathbf{UH}^{(k)},\mathbf{H}^{(k+1)})) \\
    &\mathbf{h}_\mathbf{H}^{(k+1)}, \mathbf{c}_\mathbf{H}^{(k+1)} = {\rm ConvLSTM}(\mathbf{MH}^{(k+1)},\mathbf{h}_\mathbf{H}^{(k)},\mathbf{c}_\mathbf{H}^{(k)}) \\
    & \mathbf{H}_m^{(k+1)} = {\rm CRB}(\mathbf{h}_\mathbf{H}^{(k+1)})
\end{align} 
where ${\rm ConvLSTM}$ performs similar functions as aforementioned. The features $\mathbf{KH}^{(k)}$, $\mathbf{UH}^{(k)}$ and $\mathbf{H}^{(k+1)}$ are obtained by different locations, thus possessing more adequate information and alleviate the information loss. Finally, with the output of ${\rm CNL}$ module $\mathbf{N}^{(k)}$, the updated $\mathbf{V}^{(k)}$, $\mathbf{U}^{(k)}$ and the accumulated memory state $\mathbf{H}_m^{(k)}$, we can obtain the updated $\mathbf{H}^{(k+1)}$ as illustrated in Fig.~\ref{sonfig} (c).

In summary, the signal flow of the proposed memory-augmented deep unfolding network (MADUNet) is 
\begin{align}
\setcounter{MaxMatrixCols}{12}
\begin{matrix}
  \mathbf{H}_{m}^{(k-1)} &    &\mathbf{U}_{m}^{(k-1)} &  &\mathbf{H}_{m}^{(k)}  &  &\mathbf{V}_{m}^{(k-1)}  &  &  &  &\mathbf{H}_{m}^{(k-1)} \\
       \Downarrow       &     & \Downarrow &   &\Downarrow  &  & \Downarrow &   &  & &\Downarrow \\
  \mathbf{H}^{(k)}     &\Rightarrow &  \mathbf{U}^{(k)}  &\Rightarrow   &\mathbf{H}^{(k+1)}  &\Leftarrow   &\mathbf{V}^{(k)}  &\Leftarrow &\mathbf{N}^{(k)} &\Leftarrow   &\mathbf{H}^{(k)} \\
 \Downarrow       &     & \Downarrow &   &\Downarrow  &  & \Downarrow &  &  &  &\Downarrow \\
  \mathbf{H}_{m}^{(k)} &    &\mathbf{U}_{m}^{(k)} &  &\mathbf{H}_{m}^{(k+1)}  &  &\mathbf{V}_{m}^{(k)}  &  & & &\mathbf{H}_{m}^{(k)}\\ \nonumber
\end{matrix}
\end{align}

\subsection{Network Training} \label{DUN}
The training loss for each training pair is defined as the distance between the estimated HR target image from our proposed MADUNet and the ground truth HR target image. The most widely used loss function to compute the distance is mean squared error (MSE) loss. However, MSE loss usually generates over-smoothed results. Therefore, we adopt the mean absolute error (MAE) loss to construct our training objective function, which is defined as
\begin{equation}\label{LossFunction}
 \mathcal{L} =\sum_{i=1}^{N}\left \| \mathbf{H}_{i}^{(K+1)} - \mathbf{H}_{gt,i}  \right \|_{1},
\end{equation}
where $N$ is the number of training pairs, $\mathbf{H}_{i}^{(K+1)}$ denote the $i$-th estimated HR target image, and $\mathbf{H}_{gt,i}$ is the $i$-th ground truth HR target image.

\section{Experiments}\label{sec:exp}
\vspace{-.2cm}
In this section, we evaluate the effectiveness and superiority of our proposed MADUNet on three typical GISR tasks, i.e., Pan-sharpening (section~\ref{exp:pan}), Depth Image SR (section~\ref{exp:depth}) and MR Image SR (section~\ref{exp:mri}) tasks. In each section, the specific experimental settings and  datasets are firstly described. Then, the quantitative and qualitative experimental results are reported compared with the widely-recognized state-of-the-art methods. Additionally, we conduct an ablation study to gain insight into the respective contributions of the devised components.

\begin{table*}[ht]
	\centering
	\normalsize
	\renewcommand{\tabcolsep}{3pt} % adjust horizontal space
    \renewcommand{\arraystretch}{1.2}
	\caption{\label{t1} The average quantitative results on the WorldView-II dataset (boldface highlights the best)}
	\begin{tabular}{cccccccccc}
		\toprule
		Methods & PSNR $\uparrow$& SSIM $\uparrow$ &SAM $\downarrow$ & ERGAS $\downarrow$  & SCC $\uparrow$ & Q $\uparrow$  \\
		\midrule
		SFIM ~\citep{SFIM}    &34.1297& 0.8975& 0.0439& 2.3449& 0.9079& 0.6064\\
		GS  ~\citep{GS}     & 35.6376& 0.9176& 0.0423& 1.8774& 0.9225& 0.6307 \\
		Brovey ~\citep{Brovey} & 35.8646& 0.9216& 0.0403& 1.8238& 0.8913& 0.6163\\
		IHS ~\citep{IHS}   & 35.2962& 0.9027& 0.0461& 2.0278& 0.8534& 0.5704\\
		GFPCA ~\citep{GFPCA}    & 34.558& 0.9038& 0.0488& 2.1401& 0.8924& 0.4665\\ 
		PNN  ~\citep{masi2016pansharpening} &40.7550 & 0.9624 & 0.0259 & 1.0646 & 0.9677 & 0.7426  \\
		PANNet ~\citep{yang2017pannet}  & 40.8176& 0.9626& 0.0257& 1.0557& 0.968& 0.7437\\
		MSDCNN ~\citep{yuan2018multiscale} &41.3355& 0.9664& 0.0242& 0.9940& 0.9721& 0.7577\\
		SRPPNN ~\citep{cai2020super} &41.4538& 0.9679& 0.0233& 0.9899& 0.9729& 0.7691\\
		GPPNN ~\citep{Xu_2021_CVPR} &41.1622& 0.9684& 0.0244& 1.0315& 0.9722& 0.7627\\ 
		Ours & \textbf{41.8577} & \textbf{0.9697} & \textbf{0.0229} & \textbf{0.9420}  & \textbf{0.9745}  & \textbf{0.7740} \\
		\bottomrule
	\end{tabular}%
\end{table*}%

\subsection{Pan-sharpening}\label{exp:pan}
% \vspace{-0.3cm}
To verify the effectiveness of our proposed method on the Pan-sharpening task, we conduct several experiments on the benchmark datasets compared with several representative pan-sharpening methods: 1) five commonly-recognized state-of-the-art deep-learning based methods, including PNN ~\citep{masi2016pansharpening}, PANNET ~\citep{yang2017pannet}, multiscale and multidepth network (MSDCNN) ~\citep{yuan2018multiscale}, super-resolution-guided progressive network (SRPPNN) \cite{cai2020super}, and deep gradient projection network (GPPNN) ~\citep{Xu_2021_CVPR}; 2) five promising traditional methods, including smoothing  filter-based  intensity  modulation (SFIM) ~\citep{SFIM}, Brovey ~\citep{Brovey}, GS ~\citep{GS}, intensity hue-saturation fusion (IHS) ~\citep{IHS}, and PCA  guided  filter (GFPCA) ~\citep{GFPCA}.

\subsubsection{Datasets and Evaluation Metrics}   \label{sec:dataset1}
Due to the unavailability of ground-truth pan-sharpened images, we employ the Wald protocol tool ~\citep{gt} to generate the training set. Specifically, given the MS image $\mathbf{H} \in R^{M \times N \times C}$ and the PAN image $\mathbf{P} \in R^{rM \times rN}$, both of them are down-sampled with ratio $r$, and then are denoted by $\mathbf{L} \in R^{M/r \times N/r \times C}$ and $\mathbf{p} \in R^{M \times N}$, respectively. Then, $\mathbf{L}$ is regarded as the LR MS image, $\mathbf{p}$ is the guided PAN image, and $\mathbf{H}$ is the ground truth HR MS image. In this experiment, the remote sensing images from three satellites, i.e., WorldViewII, WorldViewIII, and GaoFen2, are used for evaluation. Each database contains hundreds of image pairs, and they are divided into training, validation and testing set by  $7:2:1$.  In the training set, each training pair contains one PAN image with the size of $128 \times 128$, one LR MS patch with the size of $32 \times 32 \times 4$, and one ground truth HR MS patch with the size of $128 \times 128 \times 4$. For numerical stability, each patch is normalized to $[0,1]$.

To assess the performance of all the methods on the test data with the ground truth, we use the following image quality assessment (IQA) metrics: the relative dimensionless global error in synthesis (ERGAS), the peak signal-to-noise ratio (PSNR), the structural similarity (SSIM), the correlation coefficient (SCC), and the Q index~\citep{vivone2014critical}.

To evaluate the generalization ability of our method, we create an additional real-world full-resolution dataset of 200 samples over the newly-selected GaoFen2 satellite for evaluation. To be specific, the additional dataset is generated by the full-resolution setting where the PAN and MS images are generated as aforementioned manner without performing the down-sampling, thus PAN image is with the size of $32 \times 32$ and the MS image with the size of $128 \times 128 \times 4$. Due to the unavailability of ground-truth MS image, we adopt three commonly-used IQA metrics for assessment, i.e., the spectral distortion index $D_\lambda$, the spatial distortion index $D_S$, and the quality without reference (QNR).

\begin{figure*}
\begin{center}
\includegraphics[width=\textwidth]{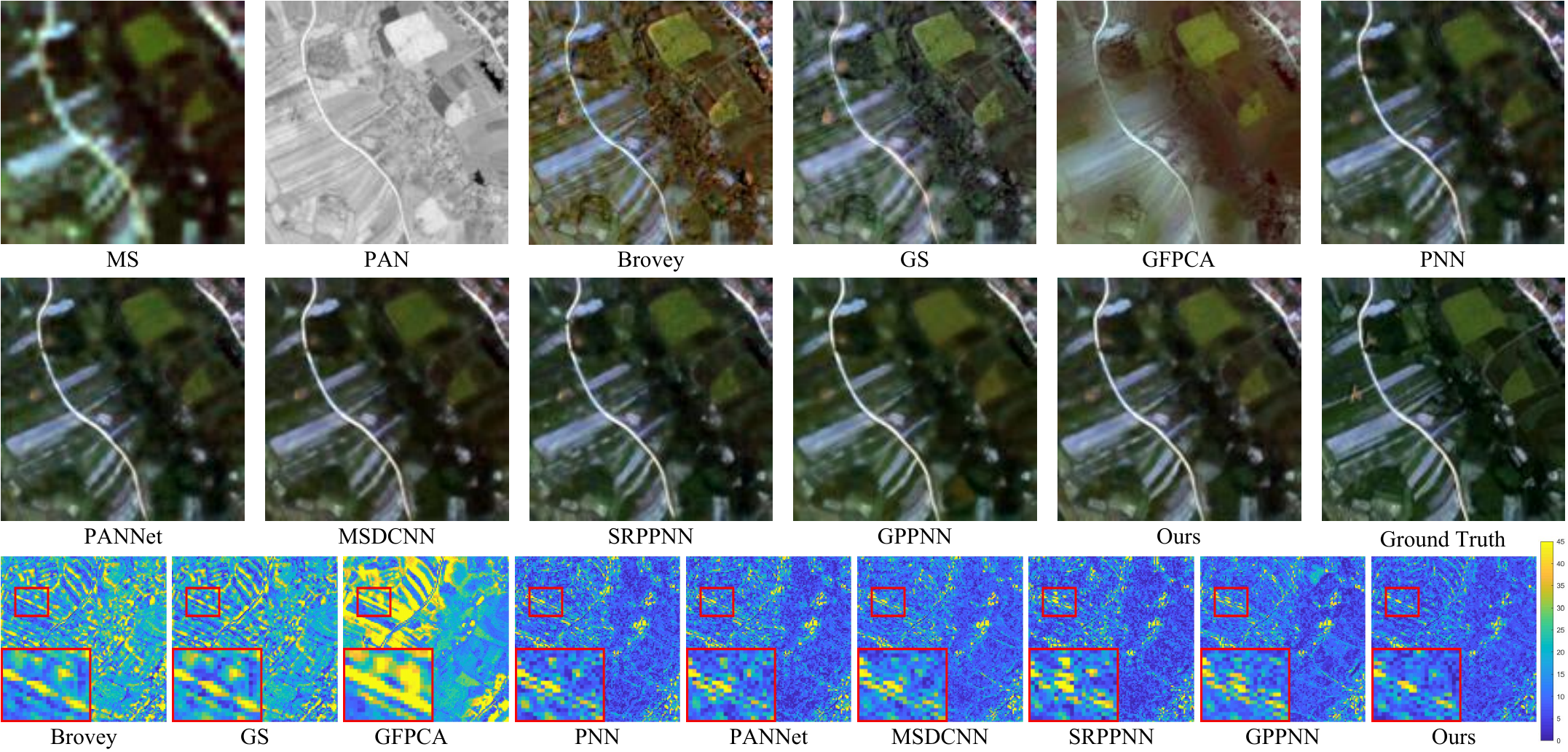}
\caption{Visual comparisons of the fused HR MS image for all the methods on one WorldView-II dataset. Images  in  the  last  row visualizes the MSE between the pan-sharpened results and the ground truth.}
\label{wv2}
\end{center}
\end{figure*}

\begin{table*}[ht]
	\centering
	\normalsize
	\renewcommand{\tabcolsep}{3pt} % adjust horizontal space
    \renewcommand{\arraystretch}{1.2}
		\caption{\label{t2}The average quantitative results on the GaoFen2 dataset (boldface highlights the best)}
	\begin{tabular}{ccccccc}
		\toprule
		Methods & PSNR $\uparrow$& SSIM $\uparrow$ &SAM $\downarrow$ & ERGAS $\downarrow$  & SCC $\uparrow$ & Q $\uparrow$ \\
		\midrule
		SFIM \citep{SFIM}   &36.906& 0.8882& 0.0318& 1.7398& 0.8128& 0.4349\\
		GS \citep{GS}    & 37.226& 0.9034& 0.0309& 1.6736& 0.7851& 0.4211\\
		Brovey \citep{Brovey}  & 37.7974& 0.9026& 0.0218& 1.3720& 0.6446& 0.3857\\
		IHS \citep{IHS}  & 38.1754& 0.91& 0.0243& 1.5336& 0.6738& 0.3682\\
		GFPCA \citep{GFPCA}   & 37.9443& 0.9204& 0.0314& 1.5604& 0.8032& 0.3236\\ 
	
		PNN \citep{masi2016pansharpening}  &43.1208& 0.9704& 0.0172& 0.8528& 0.9400& 0.739\\
		PANNet \citep{yang2017pannet}  & 43.0659& 0.9685& 0.0178& 0.8577& 0.9402& 0.7309\\
		MSDCNN \citep{yuan2018multiscale} &45.6874& 0.9827& 0.0135& 0.6389& 0.9526& 0.7759\\
		SRPPNN \citep{cai2020super} &47.1998& 0.9877& 0.0106& 0.5586& 0.9564& 0.7900\\
		GPPNN \citep{Xu_2021_CVPR} &44.2145& 0.9815& 0.0137& 0.7361& 0.9510& 0.7721\\ 
		
		Ours& \textbf{47.2668} & \textbf{0.9890} & \textbf{0.0102} & \textbf{0.5472}  & \textbf{0.9597}  & \textbf{0.7973} \\
		\bottomrule
	\end{tabular}
\end{table*}%

\begin{figure*}
\begin{center}
\includegraphics[width=\textwidth]{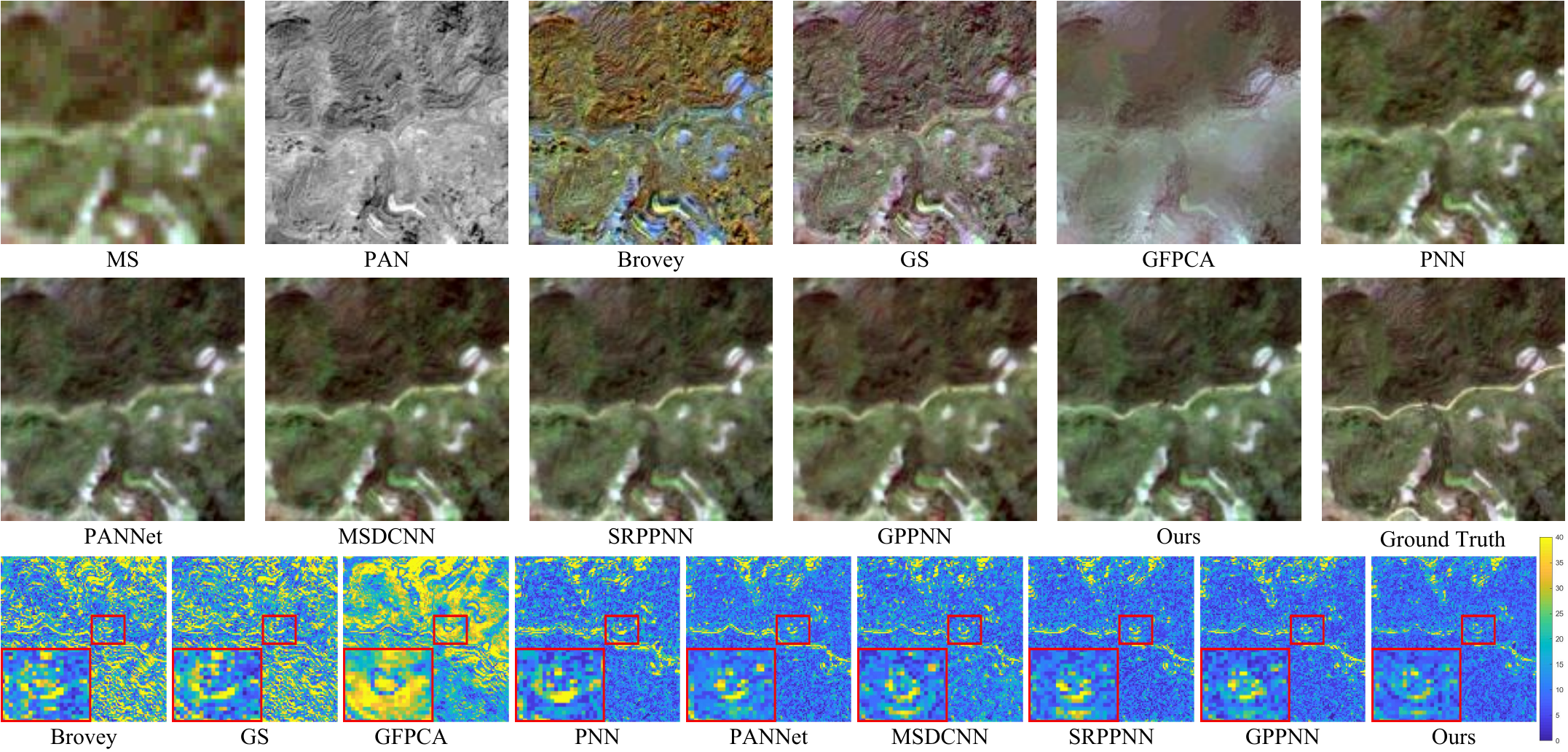}
\caption{Visual comparisons of the fused HRMS image for all the methods on one GaoFen2  dataset.  Images  in  the  last  row visualizes the MSE between the pan-sharpened results and the ground truth.}
\label{gf2}
\end{center}
\end{figure*}

\begin{figure*}
\begin{center}
\includegraphics[width=\textwidth]{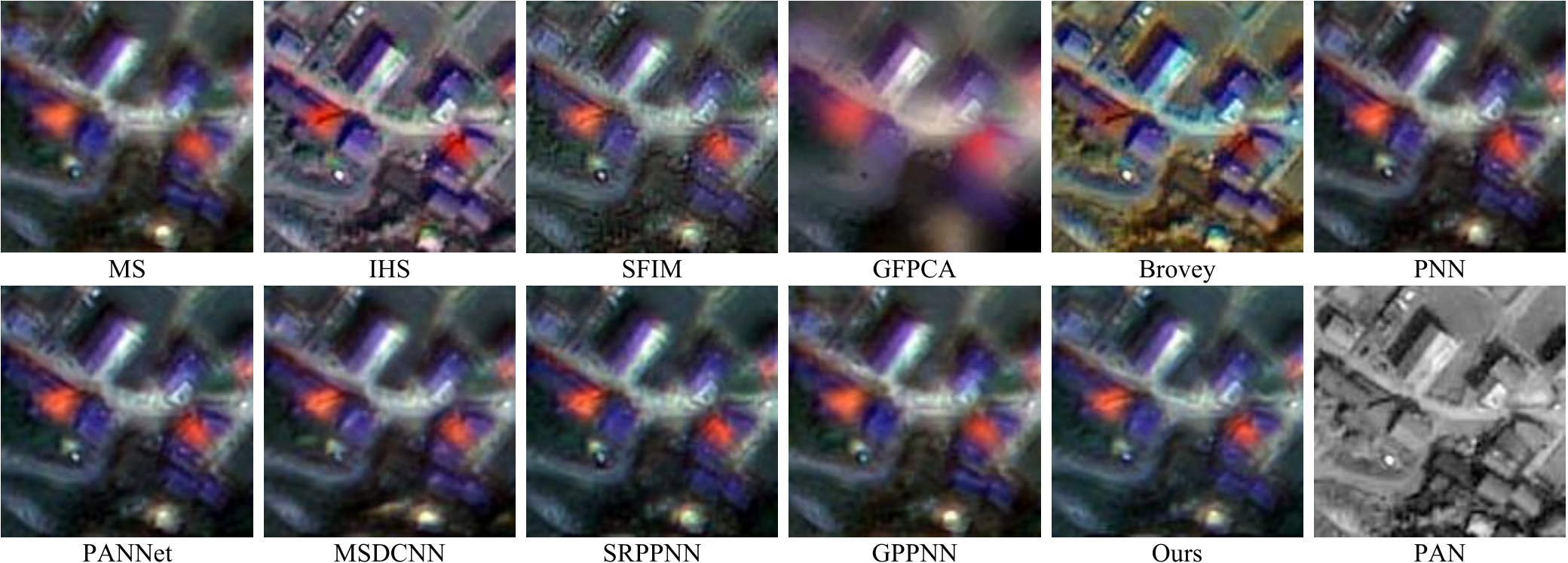}
\caption{Visual comparisons of the fused HRMS image for all the methods on a full resolution sample.}
\label{img-full}
\end{center}
\end{figure*}

\begin{table*}[t]
\normalsize
\centering
\caption{The average quantitative results on the GaoFen2 datasets in the full resolution case (boldface highlights the best)}
\resizebox{\linewidth}{!}{
\begin{tabular}{cccccccccccc}
\toprule
 Metrics& SFIM& GS& Brovey& IHS & GFPCA & PNN & PANNET & MSDCNN & SRPPNN & GPPNN & \textbf{Ours} \\ \midrule
$D_\lambda$$\downarrow$ & 0.0822 & 0.0696 & 0.1378& 0.0770 & 0.0914 & 0.0746& 0.0737 & 0.0734 & 0.0767 & 0.0782 & \textbf{0.0695} \\
$D_s$$\downarrow$ & \textbf{0.1087}& 0.2456 & 0.2605 & 0.2985& 0.1635& 0.1164 & 0.1224 & 0.1151 & 0.1162 & 0.1253 & 0.1139 \\
QNR$\uparrow$& 0.8214 & 0.7025& 0.6390& 0.6485& 0.7615& 0.8191& 0.8143 & 0.8251 & 0.8173 & 0.8073 & \textbf{0.8235} \\
\bottomrule
\end{tabular}}
\label{tab-full}
\end{table*}

\subsubsection{Implementation Details}\label{exp:detail}
In our experiments, all our designed networks are implemented in PyTorch~\citep{2019PyTorch} framework and trained on the PC with a single NVIDIA GeForce GTX 3060Ti GPU. In the training phase, these networks are optimized by the Adam optimizer~\citep{kingma2017adam} over 1000 epochs with a mini-batch size of 4. The learning rate is initialized  with $8 \times 10^{-4}$. When reaching 200 epochs, the learning rate is decayed by multiplying 0.5. Furthermore, all the hidden and cell states of ${\rm ConvLSTM}$ are initialized as zero and the input $\mathbf{H}^{(0)}$ of our unfolding network is obtained by applying Bibubic up-sampling over LR target image $\mathbf{L}$.

\subsubsection{Comparison with SOTA methods}
In this section, we will perform the  detailed quantitative and qualitative experimental analysis over datasets from WorldView-II and GaoFen2 satellites to demonstrate the effectiveness of our proposed method.

\textbf{WorldView-II Dataset Results.} Table~\ref{t1} presents the average quantitative performance between our method and aforementioned competitive algorithms over the WorldView-II dataset. From Table~\ref{t1}, it can be observed that our proposed method can significantly outperform other state-of-the-art competing methods in terms of all the metrics. To summarize at a high level, it is clearly figured out that deep learning based methods surpass the traditional methods, attributing to the powerful learning capability of deep neural networks. In particular, compared with PNN, our method is higher by 1.1027 dB and 0.0073 in PSNR and SSIM, and lower by 0.003, 0.1226 in SAM, and ERGAS, indicating better spatial information enhancement and lower spectral distortion. Compared with SRPPNN ~\citep{cai2020super}, our method performs better than SRPPNN with a large margin by 0.4039 dB on PSNR and 0.0018 on SSIM. As for the remaining metrics, our method also achieves better results than SRPPNN. Deepening into the definition of all the metrics, the best performance of our method demonstrates that our proposed method is capable of preserving precise spatial structures and avoiding the spectral distortion. In addition, we also present the visual results comparison of all the methods on one representative sample from WorldView-II dataset in Fig.~\ref{wv2}. To highlight the differences in detail, we select the R, G, B bands of the generated MS images to better visualize the qualitative comparison. As can be seen, our method can obtain better visual effect since it accurately enhance the spatial details and preserve the spectral information, which is consistent with quantitative results shown in Table~\ref{t1}. Also, the MSE residual between the generated MS image and the ground truth HR MS image are presented in the last row of Fig.~\ref{wv2}, from which we can see that traditional methods, i.e., SFIM, Brovey, and GS, suffer from notable artifacts, while the residual map of our method contains fewer details than other methods, which further verifies the superiority of our network.

\textbf{Gaofen2 Dataset Results.} Table \ref{t2} summarizes the average quantitative results of all the methods on the Gaofen2 dataset. Clearly, our proposed method outperforms other competing methods in terms of all the indexes. We can draw a similar conclusion with the Worldview II dataset. For example, deep-learning-based algorithms performs better than classical methods by a large margin. It can be seen from Table \ref{t2} that our method surpasses the second best SRPPNN~\citep{cai2020super} by 0.067 dB, 0.0013 in PSNR and SSIM, and is lower by 0.0004, 0.0114 in SAM and ERGAS. To make a visual comparison, Fig.~\ref{gf2} presents all generated pan-sharpened images and the MSE residual map between the pan-sharpened images and the ground truth images. As can be seen, our method can generate better visual images due to its better ability of enhancing the spatial details and preserving the spectral information. This visual result is consistent with the quantitative results shown in Table~\ref{t2}. For easy comparison, we also draw the MSE residual map between the generated MS image and the ground truth HR MS image in the last row of Fig.\ref{gf2}, and one small area of the residual map is amplified, from which we can observe that our method contains less structure and detail information, further verifying the superiority of our method.

\subsubsection{Effect on full-resolution scenes}
To assess the performance of our network in the full resolution case and the model generalization ability, we apply a pre-trained model built on GaoFen2 data to some unseen GaoFen2 satellite datasets with PAN of $32\times 32$ and MS of $128 \times 128 \times 8$ resolution samples. Specifically, the additional datasets are constructed using the full-resolution setting, in which the PAN and MS images are generated in their original scale using the same methods as in the preceding Section \ref{sec:dataset1}, but down-sampled. The experimental results of the all the methods are summarized in Table \ref{tab-full}. From Table \ref{tab-full}, we can observe that our proposed method performs almost the best in terms of all the indexes, which indicates that our method has better generalization ability compared with other traditional and deep learning-based methods. This is due to that our method is a model-driven deep learning method, which embeds the prior knowledge of the data into the network design and thus the proposed network naturally inherits some generalization ability of the model-based methods. Additionally, we also show visual comparisons for all the methods on a full-resolution sample in Fig.~\ref{img-full}, from which we can observe that our proposed network obtains better visual fused effect both spatially and spectrally than other competing approaches.   

\begin{figure}[]
\centering
\includegraphics[width=0.95\linewidth]{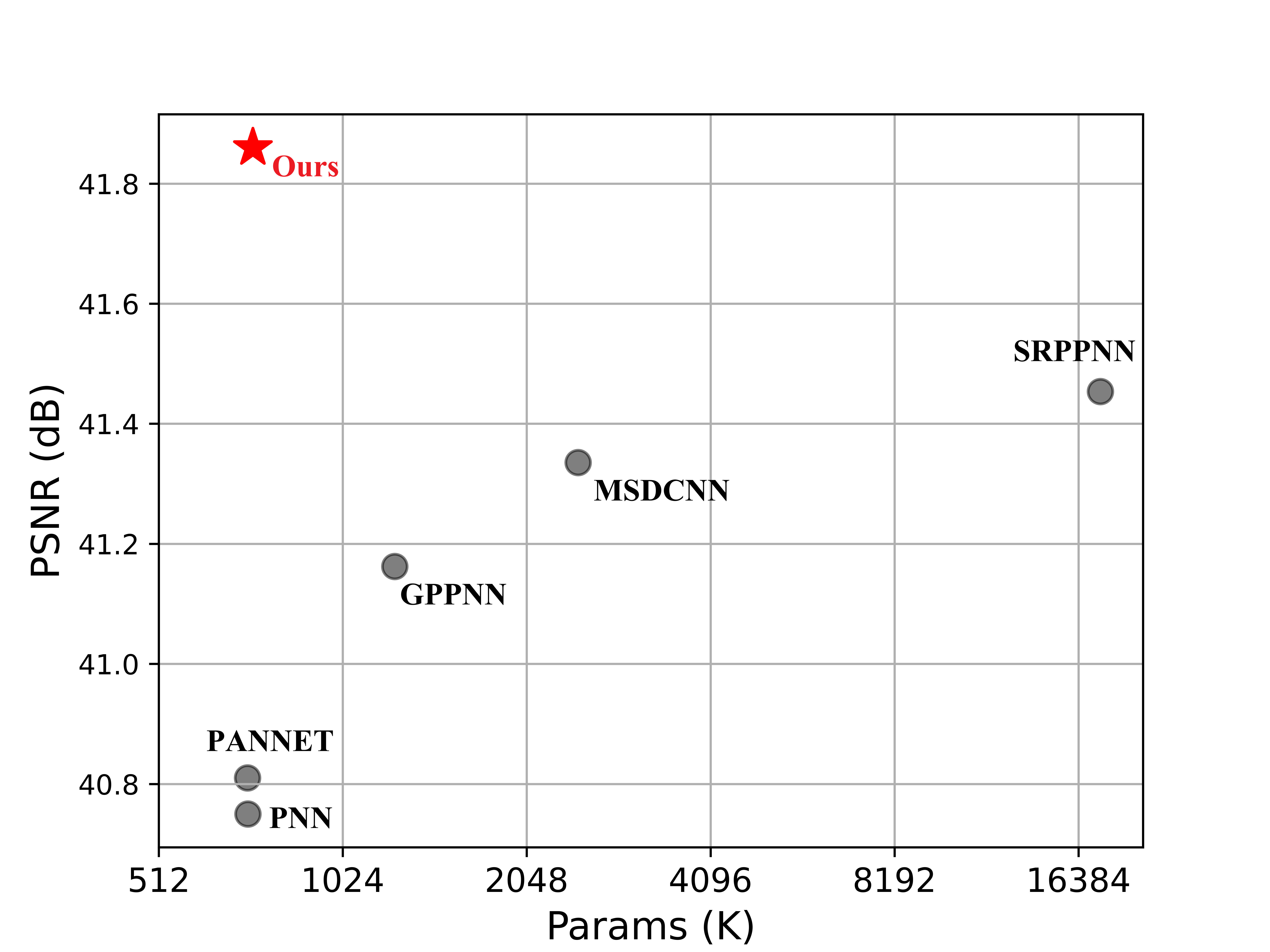}
\caption{Comparisons of model performance and number of parameters}
\label{fig:timegraph}
\vspace{-.1cm}
\end{figure}

\begin{table}[]
\caption{Comparisons on parameter numbers, FLOPs and time.}
 \label{tab:pparams}
\footnotesize
\renewcommand{\tabcolsep}{2pt} % adjust horizontal space
\renewcommand{\arraystretch}{1.2} % adjust vertical space
\centering
\begin{tabular}{ccccccc}
\toprule
  Methods     & PNN & PANNet & MSDCNN & SRPPNN & GPPNN & Ours \\ \midrule
$\#$Params & 0.689    &  0.688      &  2.390      &   17.114    & 1.198     & 0.700     \\
FLOPs  &  1.1289   &   1.1275     &  3.9158      &   21.1059     &  1.3967     & 4.4543    \\ 
Time (s) & 0.0028   &  0.0027      &  0.0037      &   0.0098    & 0.0075     & 0.0081    \\ \bottomrule
\end{tabular}
\end{table}

\subsubsection{Complexity Analysis}
To conduct a more thorough analysis of the methods, we investigate the complexity of the proposed method, including the  floating-point operations (FLOPs), the number of parameters (in 10 M) and running time in this section.  Comparisons on parameter numbers and model performance (as measured by PSNR) are provided in Table \ref{tab:pparams} and Fig.~\ref{fig:timegraph}.  As can be seen, PNN~\citep{masi2016pansharpening} and PANNet~\citep{yang2017pannet} have the fewest FLOPs due to their network architectures containing only a few convolutional layers. Additionally, they extract a small number of feature maps, resulting in a smaller FLOPs count. MSDCNN~\citep{yuan2018multiscale} and SRPPNN~\citep{cai2020super} both exhibit large increases in the number of parameters and FLOPs as the number of convolutional layers and the complexity of the network design rise. Notably, they also achieve the second and third highest performance at the expense of massive model computation and storage. Additionally, the most comparable solution to ours, GPPNN~\citep{Xu_2021_CVPR}, is organized around the model-based unfolding principle and has comparable model parameters and flops reductions but inferior performance. This is due to powerful model learning's incapability without fully exploring the potential of different modalities. Also, our method is comparable with other methods in terms of running time. In summary, our network achieves a favorable trade-off between calculation, storage, and model performance compared with other methods.

\begin{table*}[!tb]
\setlength{\tabcolsep}{7pt}%\renewcommand{\arraystretch}{1.2}
\centering
\renewcommand\arraystretch{1.2}
	\begin{center}
	\caption{\label{tab:nyu_result_bic} Average RMSE performance comparison for scale factors $4 \times, 8 \times$ and $16 \times$ with bicubic down-sampling. The best performance is shown in \textbf{bold} and second best performance is underlined.}
		\begin{tabular}{l|rrr|rrr|rrr|rrr}
		\toprule
		\multirow{2}{*}{Method} & \multicolumn{3}{c|}{Middlebury} & \multicolumn{3}{c|}{Lu} & \multicolumn{3}{c|}{NYU v2}  & \multicolumn{3}{c}{Average} \\
		\cline{2-13}
		~ &$4\times$ & $8 \times$  & $16\times$  & $4\times$ & $8 \times$  & $16\times$  &$4\times$ & $8 \times$  & $16\times$ &$4 \times$ & $8 \times$  & $16\times$ \\
		\hline
		Bicubic & 2.47 & 4.65 & 7.49 &  2.63  & 5.23 & 8.77 & 4.71 & 8.29 & 13.17 & 3.27 & 6.06& 9.81   \\
		GF \citep{GF} & 3.24 & 4.36 & 6.79 & 4.18 & 5.34 & 8.02 & 5.84 & 7.86 & 12.41 & 4.42 & 5.85 & 9.07\\		
		TGV \citep{TGV} & 1.87 & 6.23 & 17.01 & 1.98 & 6.71 & 18.31 & 3.64 & 10.97 & 39.74 & 2.50 & 7.97 & 25.02 \\
		DGF~\citep{DGF}  & 1.94 & 3.36 & 5.81 & 2.45 & 4.42 & 7.26 & 3.21 & 5.92 & 10.45 & 2.53 & 4.57 &  7.84 \\
		DJF~\citep{DJF} & 1.68 & 3.24 & 5.62 & 1.65 & 3.96 & 6.75 & 2.80 & 5.33 & 9.46 & 2.04 & 4.18 & 7.28 \\
		DMSG~\citep{DMSG} & 1.88 & 3.45 & 6.28 & 2.30 & 4.17 & 7.22 & 3.02 & 5.38 & 9.17 & 2.40 & 4.33 & 7.17  \\
		DJFR~\citep{DJFR}& 1.32 & 3.19 & 5.57 & 1.15 & 3.57 & 6.77 & 2.38 & 4.94 & 9.18  & 1.62 & 3.90  & 7.17\\
		DSRNet~\citep{DSRnet} & 1.77 &  3.05 &  4.96 & 1.77 &  3.10 &  \underline{5.11} & 3.00 &  5.16 &  8.41 & 2.18  & 3.77 & 6.16 \\
		PacNet~\citep{PacNet} &  1.32 & 2.62 & 4.58 & 1.20 & 2.33 & 5.19 & 1.89 & 3.33 & 6.78& 1.47  & 2.76 & 5.53 \\
		FDKN~\citep{DKN} & \underline{1.08} & 2.17 &  4.50 & \textbf{0.82} & 2.10 & 5.05 & 1.86 & 3.58 & 6.96 & 1.25 & 2.62 & 5.50 \\
		DKN~\citep{DKN} &  1.23 & \underline{2.12} &\underline{4.24} & 0.96 & \underline{2.16} & \underline{5.11} & \underline{1.62} & \underline{3.26} & \underline{6.51} & 1.27 & 2.51 &5.29 \\
		Ours & \textbf{1.15} & \textbf{1.69} & \textbf{3.23} & \underline{0.90}& \textbf{1.74} &\textbf{3.86} & \textbf{1.51} & \textbf{3.02} & \textbf{6.23} & \textbf{1.18} & \textbf{2.15} & \textbf{4.44} \\
		\bottomrule
		\end{tabular}
	\end{center}
\end{table*}

\subsection{Depth Image SR}~\label{exp:depth}
In this section, to verify the effectiveness of our method, we conduct several experiments to compare our method with the representative state-of-the-art Depth Image SR methods. Following the experimental protocol of ~\citep{DKN}, we generate the evaluated datasets over two down-samping operations, i.e., bibubic down-sampling and direct down-sampling. The detailed analysis is described as bellow.

\subsubsection{Datasets and Metrics}
\label{dataset}
 \textbf{Dataset.} NYU v2 dataset~\citep{NYU} is the widely-recognized benchmark for Depth Image SR. This dataset consists of 1449 RGB-D image pairs recorded by Microsoft Kinect sensors in the structured light. Following the same settings as previous Depth Image SR methods~\citep{DJFR, DKN}, we train our proposed network over the first 1000 RGB-D image pairs and then evaluate the trained model on the remaining 449 RGB-D images pairs. We follow the experimental protocol of ~\citep{DKN} to build the low-resolution depth map, which involves using the bibubic and direct down-sampling operations at different ratios ($\times 4$, $\times 8$ and $\times 16$), respectively. In addition, the root mean squared error (RMSE) is adopted by default to evaluate the model performance. 
 
Furthermore, to determine the potential generalization ability of the model, following the same setting as ~\citep{DKN}, we directly test the trained model over NYU v2 dataset over the additional Middlebury dataset~\citep{middleblur_data_1} and  Lu~\citep{Lu} dataset, which are another two benchmark datasets to evaluate the performance of Depth Image SR algorithms. The Middlebury dataset contains 30 RGB-D image pairs, of which 21 pairs are from 2001~\citep{2001} and 9 pairs are from 2006~\citep{middleblur_data_2}. The Lu dataset contains 6 RGB-D image pairs. Following the existing works~\citep{DSRnet, PMBANet, DKN}, we quantify all the recovered depth maps to 8-bits before calculating the MAE or RMSE values for fair evaluation. For both criteria, lower values indicate higher performance. 

\begin{figure*}
\begin{center}
\includegraphics[width=\textwidth]{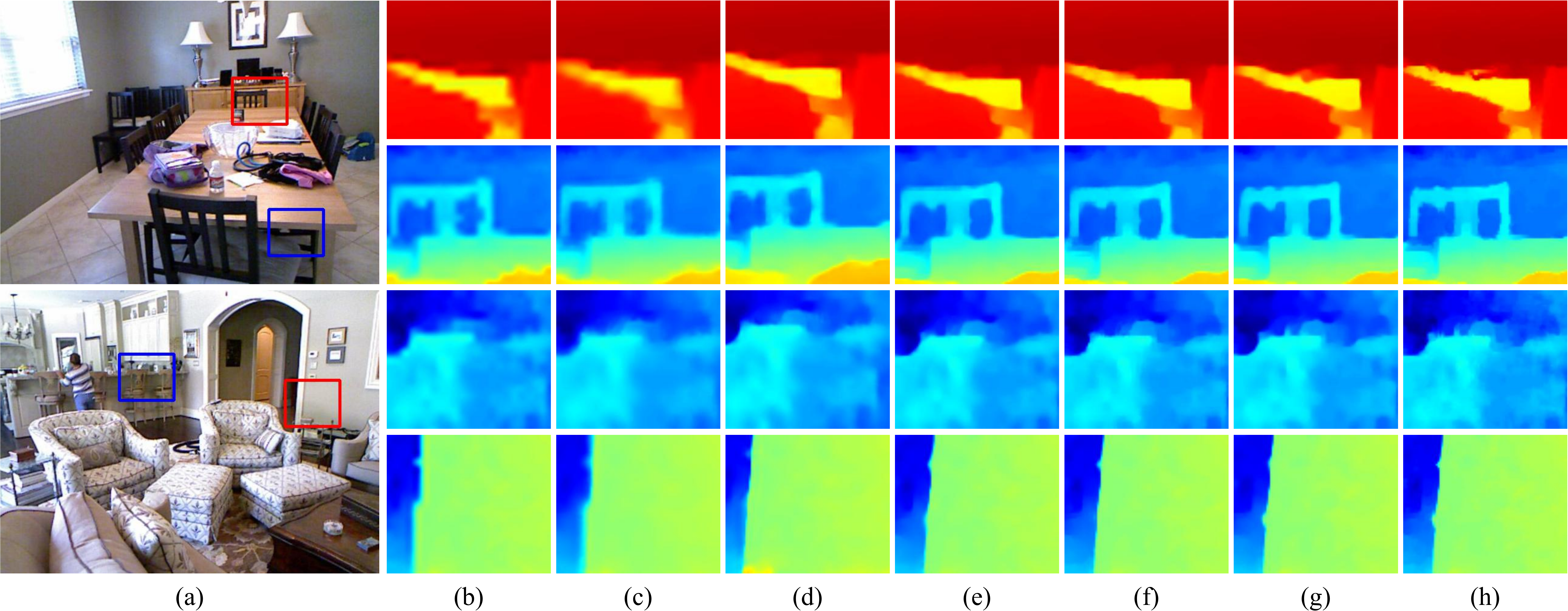}
\caption{Visual comparison of 4$\times$ upsampling results: (a) RGB image, (b) Bibubic,
(c) GF, (d) DJF, (e) PacNet, (f) DKN, (g) Ours, (h) GT.}
\label{RGBD_4x}
\end{center}
\end{figure*}

\begin{figure*}
\begin{center}
\includegraphics[width=\textwidth]{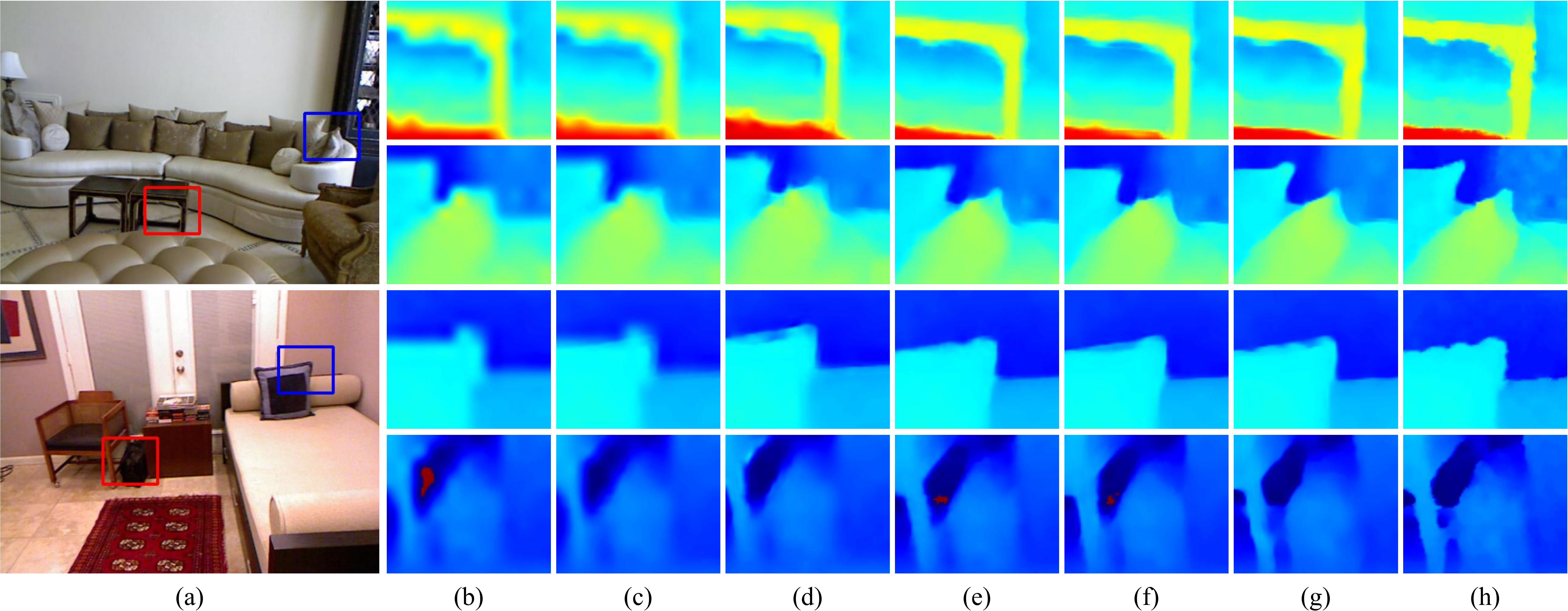}
\caption{Visual comparison of 8$\times$ upsampling results: (a) RGB image, (b) Bibubic,
(c) GF, (d) DJF, (e) PacNet, (f) DKN, (g) Ours, (h) GT.}
\label{RGBD_8x}
\end{center}
\end{figure*}

\begin{figure*}
\begin{center}
\includegraphics[width=\textwidth]{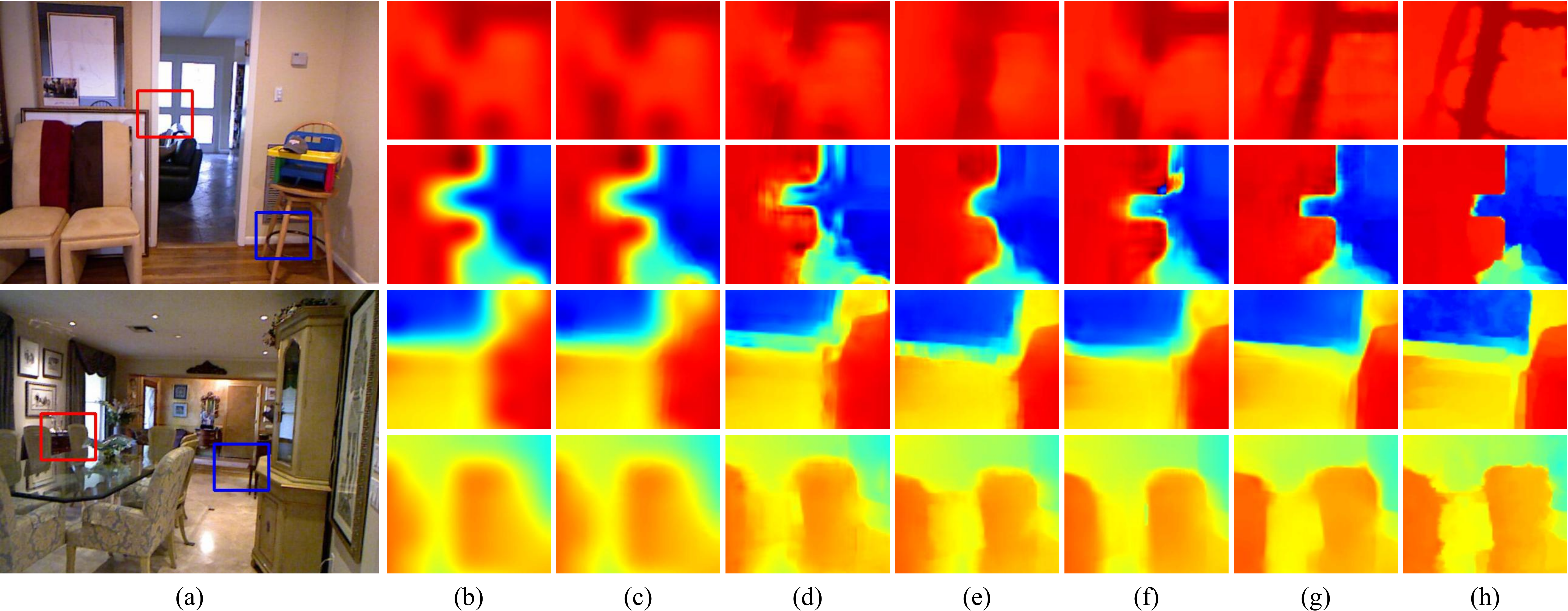}
\caption{Visual comparison of 16$\times$ upsampling results: (a): RGB image, (b) Bibubic,
(c) GF, (d) DJF, (e) PacNet, (f) DKN, (g) Ours, (h) GT.}
\label{RGBD_16x}
\end{center}
\end{figure*}

\begin{figure*}
\begin{center}
\includegraphics[width=\textwidth]{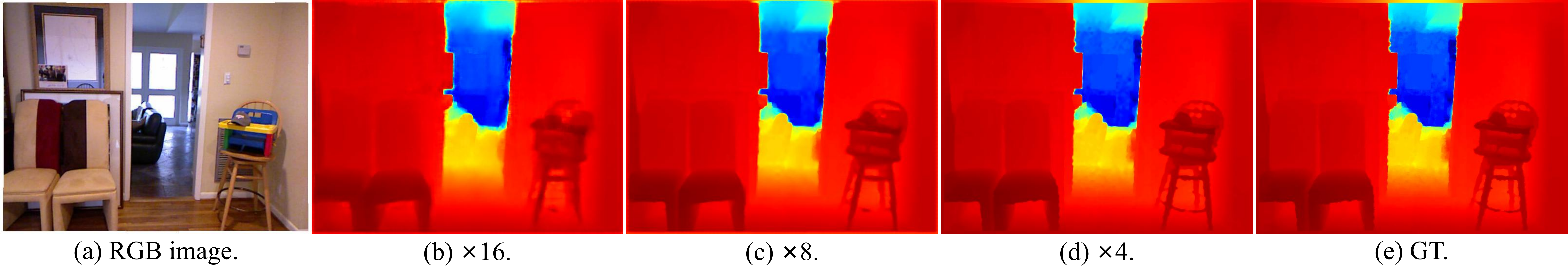}
\caption{Visual comparison between 16$\times$, 8$\times$ and 4$\times$ upsampling results of our method.}
\label{RGBD_4816}
\end{center}
\end{figure*}

\subsubsection{Implementation Details}
\label{id}
 In our experiments, all our designed networks are implemented in PyTorch~\citep{2019PyTorch} framework and trained on the PC with a single NVIDIA GeForce GTX 3060Ti GPU. In the training phase, we choose the released resource code of ~\citep{DKN} as baseline to implement all our experiments. Specifically, all the our networks are optimized the Adam optimizer~\citep{kingma2017adam} with $\beta_{1}=0.9, \beta_{2}=0.999,$ and $\epsilon=1 \mathrm{e}-8$.  The initial learning rate is $2 \times 10^{-4}$. For every 100 epochs, the learning rate is decayed by multiplying 0.5. Furthermore, all the hidden and cell states of ${\rm ConvLSTM}$ are initialized as zero and the input $\mathbf{H}^{(0)}$ of our unfolding network is obtained by applying bibubic up-sampling over LR target image $\mathbf{L}$.

\subsubsection{Experimental Results on Bibubic Down-sampling}
To evaluate the effectiveness of the proposed method on the bibubic down-sampling case, following the same setting as ~\citep{DKN}, we select the representative state-of-the-art Depth Image SR algorithms for comparison: 1) traditional methods, including guided image filtering (GF) ~\cite{GF} and total  generalized  variation (TGV) ~\citep{TGV}, 2) deep learning based methods, containing DGF~\citep{DGF}, DJF~\citep{DJF}, DMSG~\citep{DMSG}, deep joint image filte (DJFR)~\citep{DJFR}, depth  super-resolution network (DSRNet)~\citep{DSRnet}, pixel-adaptive convolution (PacNet)~\citep{PacNet}, fast deformable Kernel Networks (FDKN)~\citep{DKN} and deformable Kernel Networks (DKN)~\citep{DKN}. All the results are obtained from the original papers reported by the authors  for fair comparison. Table~\ref{tab:nyu_result_bic} reports the quantitative comparison among all the competing algorithms at different ratios, i.e., $\times 4$, $\times 8$ and $\times 16$. The average RMSE values between the generated HR depth map and the ground truth depth map are illustrated.  

From Table~\ref{tab:nyu_result_bic}, we can observe that our proposed network performs best compared with other methods at different ratios of $\times 4$, $\times 8$ and $\times 16$ by a large margin in average RMSE values. To highlight at a high level, deep-learning based methods~\citep{DKN, DSRnet, PacNet, DMSG, DJFR} achieve better results than traditional methods~\citep{GF, TGV} by significant margins in terms of RMSE, which is attributed to the powerful learning and mapping capability of deep neural networks. 
Compared with the second best method DKN~\citep{DKN}, our method decreases the average RMSE over all the three datasets by $0.09$~($4 \times$), $0.36$~($8 \times$) and $0.85$~($16 \times$), respectively. To evaluate the generalization ability, the models are well trained over NYU v2 dataset and not further fine-tuned over other datasets (i.e., Middlebury dataset and Lu dataset). From Table~\ref{tab:nyu_result_bic}, it can be seen that our method obtains superior performance on the Middlebury dataset and Lu dataset, which implies that our method has better generalization ability compared with other methods.

In terms of visual comparison, Fig.~\ref{RGBD_4x},  Fig.~\ref{RGBD_8x} and  Fig.~\ref{RGBD_16x} display all the generated latent HR depth maps. As can be seen, other competing methods exist some issues. For example, GF~\citep{GF} generates over-smoothed map since the local filter cannot capture global information. DJFR~\citep{DJFR} and DKN~\citep{DKN} suffer from diffusion artifacts. PacNet~\citep{PacNet} can preserve the local details, but cannot reconstruct the boundary well. On the contrary, our proposed method obtains the best visual effect than other competing methods since it can enhance the spatial details of LR depth maps, and generate accurate and sharp edges. 

Additionally, we also select one representative sample to verify the effect difference over different ratios, as shown in Fig.~\ref{RGBD_4816}, from which we can see that the $\times 4$ up-sampling results are more reasonable than that of $\times 8$ and $\times 16$. The reason is that the higher ratios of down-sampling loss much more information than the lower ratio case, thus resulting in that the reconstruction process to be difficult. The consistent conclusion can also be supported by the quantitative results reported in Table~\ref{tab:nyu_result_bic}.

\subsubsection{Experimental Results on Direct Down-sampling}
In terms of the direct down-sampling case, we compare the proposed method with the following state-of-the-art methods: directly Bibubic up-sampling, markov random fields (MRF)~\citep{MRF}, guided image filtering (GF)~\citep{GF}, total generalized variation network (TGV)~\citep{TGV}, high quality depth map up-sampling (Park)~\citep{Park}, joint static and dynamic guidance ~\citep{Ham}, joint bilateral up-sampling (JBU)~\citep{JBU}, deep joint image filtering (DJF)~\citep{DJF}, deep multi-scale guidance network (DMSG)~\citep{DMSG}, pixel-adaptive convolution neural network (PacNet)~\cite{PacNet}, deep joint image filter (DJFR)~\citep{DJFR}, depth super-resolution network (DSRNet)~\citep{DSRnet}, fast deformable kernel network (DKN)~\citep{DKN} and deformable kernel network (DKN)~\citep{DKN}. All the results for other competing methods are directly taken from the original paper reported by the authors for fair comparison. Table~\ref{tab:nyu_result} presents the quantitative comparison among all the competing algorithms at different ratios, i.e., $4\times$, $8\times$ and $16\times$. The average RMSE values are recorded.

\begin{table*}[!tb]\setlength{\tabcolsep}{7pt}%\renewcommand{\arraystretch}{1.2}
\centering
\renewcommand\arraystretch{1.2}
	\begin{center}
	\caption{\label{tab:nyu_result}Average RMSE performance comparison for scale factors $4 \times, 8 \times$ and $16 \times$ with direct down-sampling. The best performance is shown in \textbf{bold} and second best performance is underlined.}
		\begin{tabular}{l|rrr|rrr|rrr|rrr}
		\toprule
		\multirow{2}{*}{Method} & \multicolumn{3}{c|}{Middlebury} & \multicolumn{3}{c|}{Lu} & \multicolumn{3}{c|}{NYU v2}  & \multicolumn{3}{c}{Average} \\
		\cline{2-13}
		~ &$4\times$ & $8 \times$  & $16\times$  & $4\times$ & $8 \times$  & $16\times$  &$4\times$ & $8 \times$  & $16\times$ &$4 \times$ & $8 \times$  & $16\times$ \\
		\hline
		Bicubic & 4.44 & 7.58 & 11.87 & 5.07 & 9.22 & 14.27 & 8.16 & 14.22 & 22.32 & 5.89 & 10.34 & 16.15 \\
		MRF \citep{MRF} & 4.26 & 7.43 & 11.80 & 4.90 & 9.03 & 14.19 & 7.84 & 13.98 & 22.20 & 5.67 & 10.15 & 16.06 \\
		GF \citep{GF} & 4.01 & 7.22 & 11.70 & 4.87 & 8.85 & 14.09 & 7.32 & 13.62 & 22.03 & 5.40 & 9.90 & 15.94\\
		TGV \citep{TGV} & 3.39 & 5.41 & 12.03 & 4.48 & 7.58 & 17.46 & 6.98 & 11.23 & 28.13 & 4.95 & 8.07 & 19.21 \\
		Park \citep{Park} & 2.82 & 4.08 & 7.26 & 4.09 & 6.19 & 10.14 & 5.21 & 9.56 & 18.10 & 4.04 & 6.61 & 11.83 \\
		Ham \citep{Ham} & 3.14 & 5.03 & 8.83 & 4.65 & 7.73 & 11.52 & 5.27 & 12.31 & 19.24 & 4.35 & 8.36 &  13.20 \\
		JBU \citep{JBU} & 2.44 & 3.81 & 6.13 & 2.99 & 5.06 & 7.51 & 4.07 & 8.29 & 13.35 & 3.17 & 5.72 & 9.00\\
		DGF~\citep{DGF} & 3.92& 6.04& 10.02& 2.73& 5.98& 11.73& 4.50& 8.98& 16.77 & 3.72 & 7.00 & 12.84 \\
		DJF \citep{DJF} & 2.14 & 3.77 & 6.12 & 2.54 & 4.71 & 7.66 & 3.54 & 6.20 & 10.21 & 2.74 & 4.89 & 8.00 \\
		DMSG \citep{DMSG} & 2.11 & 3.74 & 6.03 & 2.48 & 4.74 & 7.51 & 3.37 & 6.20 & 10.05 & 2.65 & 4.89 & 7.86 \\
		PacNet~\citep{PacNet} & 1.91 & 3.20 & 5.60 & 2.48 & 4.37 & 6.60 & 2.82 & 5.01 & 8.64 & 2.40 & 4.19 & 6.95 \\
        DJFR \citep{DJFR} & 1.98 & 3.61 & 6.07 & \underline{2.21} & \textbf{3.75} & 7.53 & 3.38 & 5.86 & 10.11 & 2.52 & 4.41 & 7.90 \\
        DSRNet~\citep{DSRnet} & 2.08 & 3.26 &  5.78 &  2.57 &  4.46 &  6.45 &  3.49 &  5.70  & 9.76  & 2.71 & 4.47 & 7.30 \\
        FDKN \citep{DKN}& 2.21 & 3.64 & 6.15 & 2.64 & 4.55 & 7.20 & 2.63 & 4.99 & 8.67 & 2.49 & 4.39 & 7.34\\
		DKN \citep{DKN} & \underline{1.93} & \underline{3.17} & \underline{5.49} & 2.35 & 4.16 & \underline{6.33} & \underline{2.46} & \underline{4.76} & \underline{8.50} & \underline{2.25} & \underline{4.03} & \underline{6.77} \\
		Ours & \textbf{1.82} & \textbf{2.92} & \textbf{5.14} & \textbf{2.13} & \underline{3.88} & \textbf{6.30} & \textbf{2.37} & \textbf{4.61} & \textbf{8.32} & \textbf{2.10} & \textbf{3.80} & \textbf{6.58} \\
		\bottomrule
		\end{tabular}
	\end{center}
\end{table*}

In comparison with other approaches, our method obtains the best performance in terms of the average RMSE values for all the scaling factors. In the most difficult $\times 16$ case for all the datasets, our proposed method obviously outperforms other methods
as shown in Table~\ref{tab:nyu_result}. Specifically, compared with the second best method DKN~\citep{DKN}, our method decreases the average RMSE by $0.15$~($4 \times$), $0.23$~($8 \times$) and $0.19$~($16 \times$), respectively. Additionally, to evaluate the generalization ability, our model is trained over the NYU v2 dataset and then is directly used for testing on the other two datasets, i.e., Middlebury dataset and Lu dataset. From Table~\ref{tab:nyu_result}, we can observe that our method achieves better performance on the Middlebury dataset and Lu dataset compared with other methods, which implies that our method has  good generalization ability. In particular, compared with the second best model DKN~\citep{DKN}, our method decreases the average RMSE by $0.11$~($4 \times$), $0.25$~($8 \times$) and $0.35$~($16 \times$) on the Middlebury dataset~\citep{middleblur_data_1}, and
$0.22$~($4 \times$), $0.28$~($8 \times$) and $0.03$~($16 \times$)  on the Lu~\citep{Lu} dataset.

\subsubsection{Parameter Comparison}
To conduct a more thorough evaluation, we also investigate the complexity of our method on the parameter number. Table \ref{tab:params} records the parameter number and model performance (evaluated by RMSE index) in the ratio $\times 4$ case. As can be seen, our method obtains the best performance while contains the second fewest parameters. As for other competing methods, DJFR~\citep{DJFR} and PacNet~\citep{PacNet} have the fewest model storage due to their simple network architectures, but results in poor performance. DSRnet~\citep{DSRnet} and PMBAN~\citep{PMBANet} both exhibit large increases in the number of parameters, but they also achieve a slightly decrease on the RSME value. Additionally, DKN~\citep{DKN} achieve the second best performance but has more parameters than our method. To emphasize, our network achieves a favorable trade-off between model complexity and model performance compared with other state-of-the-art methods.

\begin{table}[]
\caption{Comparisons on parameter numbers and model performance of depth image SR at ratio $\times 4$ over NYU-v2 dataset.}
 \label{tab:params}
\footnotesize
\renewcommand{\tabcolsep}{2.1pt} % adjust horizontal space
\renewcommand{\arraystretch}{1.2} % adjust vertical space
\centering
\begin{tabular}{cccccccc}
\toprule
  Methods     & DMSG & DJFR & DSRNet & PacNet & DKN & PMBAN & Ours \\ \midrule
$\#$Params & 0.33    &  \textbf{0.08}      &  45.49      &   0.18   & 1.16  & 25.06 & 0.13   \\
RMSE  &  3.02   &   2.38    &  3.00     &   1.89    &  1.62   & 1.73   & \textbf{1.51}    \\ \bottomrule
\end{tabular}
\end{table}

\subsection{MRI Image SR}\label{exp:mri}
% \vspace{-0.3cm}
This section conducts a serious of experiments to evaluates the performance our method on the MR Image SR task.
Several representative MR Image SR methods are used for comparison, including Bibubic up-sampling, plain convolutional neural network (PCNN)~\citep{zeng2018simultaneous}, progressive network (PGN)~\citep{lyu2020multi}, multi-stage integration network (MINet)~\citep{feng2021multi}.

\subsubsection{Datasets and Implementation Details}   \label{sec:dataset}
In MR Image SR, it produces different contrast images of T1 and T2 but with the same anatomical structure. Due to the complementary property of T1 and T2,  MR Image SR aims to super-solve the low-spatial resolution T1 image with the guidance of high-resolution T2. Following the setting of ~\citep{zeng2018simultaneous}, we generates the MR Image SR datasets for two types of settings as bellow. Given the T1 and T2 images with the size of $300 \times 256 \times 1$, T1 is down-sampled into $75 \times 64 \times 1$ with ratio $4$, and $150 \times 128 \times 1$ resolutions with ratio $2$. The generated dataset is split into the training, validation and testing part by $7:2:1$ respectively, and each one contains hundreds of samples. To assess the model performance, the peak signal-to-noise ratio (PSNR) and the structural similarity (SSIM) are adopted.  

In our experiments, all our designed networks are implemented in PyTorch~\citep{2019PyTorch} framework and trained on the PC with a single NVIDIA GeForce GTX 3060Ti GPU. In the training phase, these networks are optimized by the SGD optimizer over 100 epochs with a mini-batch size of 2. Furthermore, all the hidden and cell states of ${\rm ConvLSTM}$ are initialized as zero and the input $\mathbf{H}^{(0)}$ of our unfolding network is obtained by applying Bibubic upsampling over LR T1 image $\mathbf{L}$. 

\begin{table}[t]
\footnotesize
\centering
\renewcommand{\tabcolsep}{2.3pt}
\renewcommand\arraystretch{1.2}
\caption{Average performance comparison over $\times 4$ MR Image SR.}
\begin{tabular}{cccccc}
\toprule
 Metrics & Bibubic & PCNN & PGN & MINet & \textbf{Ours} \\ \midrule
PSNR$\uparrow$ & 21.2330 & 32.9334 & 33.5145 & 35.1998 & \textbf{35.2928} \\
SSIM$\uparrow$  & 0.6773 & 0.8933 & 0.9011 & 0.9190 & \textbf{0.9221}\\
\bottomrule
\end{tabular}
\label{tab-mri}
\end{table}

\begin{table}[t]
\footnotesize
\centering
\renewcommand{\tabcolsep}{2.3pt}
\renewcommand\arraystretch{1.2}
\caption{Average performance comparison over $\times 2$ MR Image SR.}
\begin{tabular}{cccccc}
\toprule
 Metrics & Bibubic & PCNN & PGN & MINet & \textbf{Ours} \\ \midrule
PSNR$\uparrow$ & 27.3613 & 33.8824 & 35.74215 & 38.2553 & \textbf{38.3110} \\
SSIM$\uparrow$  & 0.7576 & 0.9102 & 0.9236 & 0.9449 & \textbf{0.9476}\\
\bottomrule
\end{tabular}
\label{tab-mri-2}
\end{table}

\begin{figure*}
\begin{center}
\includegraphics[width=\textwidth]{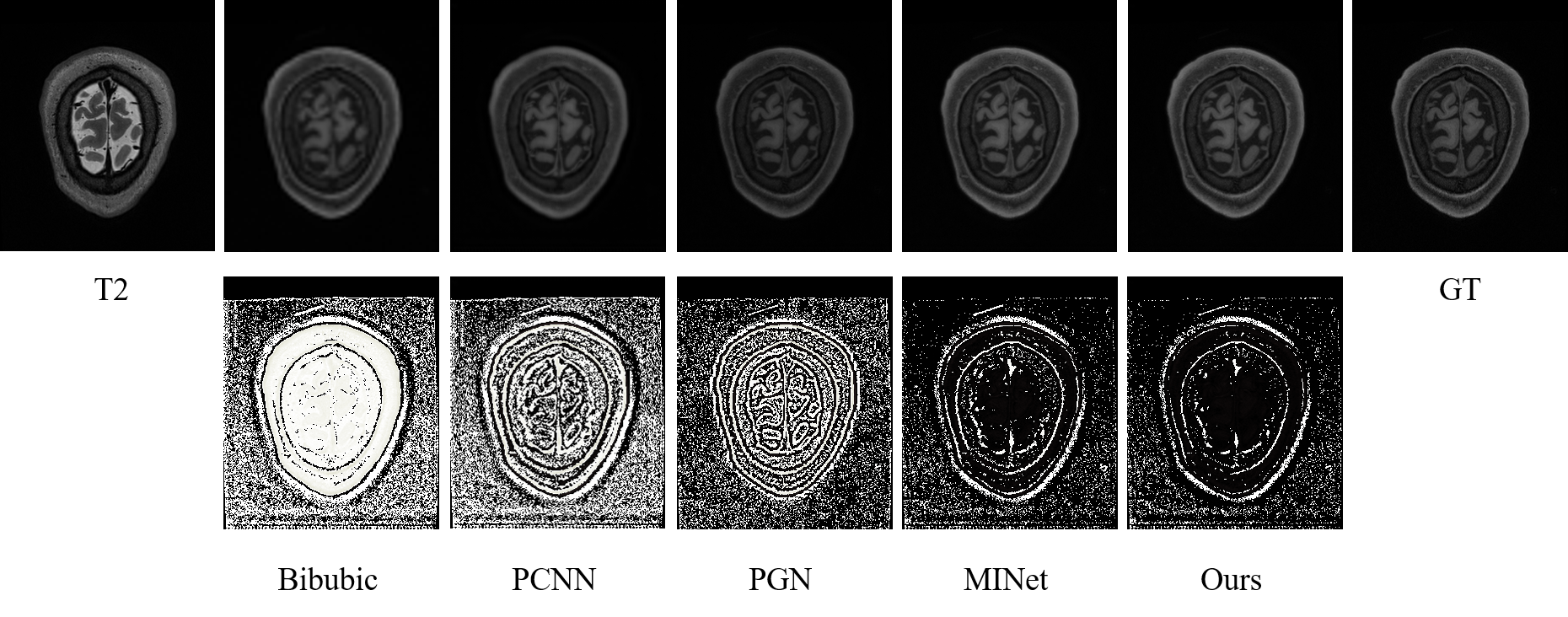}
\caption{Visual comparison of $\times 2$ MRI super-resolution and T2 denotes the guidance image.  Images in the last row visualizes the MSE between the generated results and the ground truth.}
\label{mrifig}
\end{center}
\end{figure*}

\begin{figure*}
\begin{center}
\includegraphics[width=\textwidth]{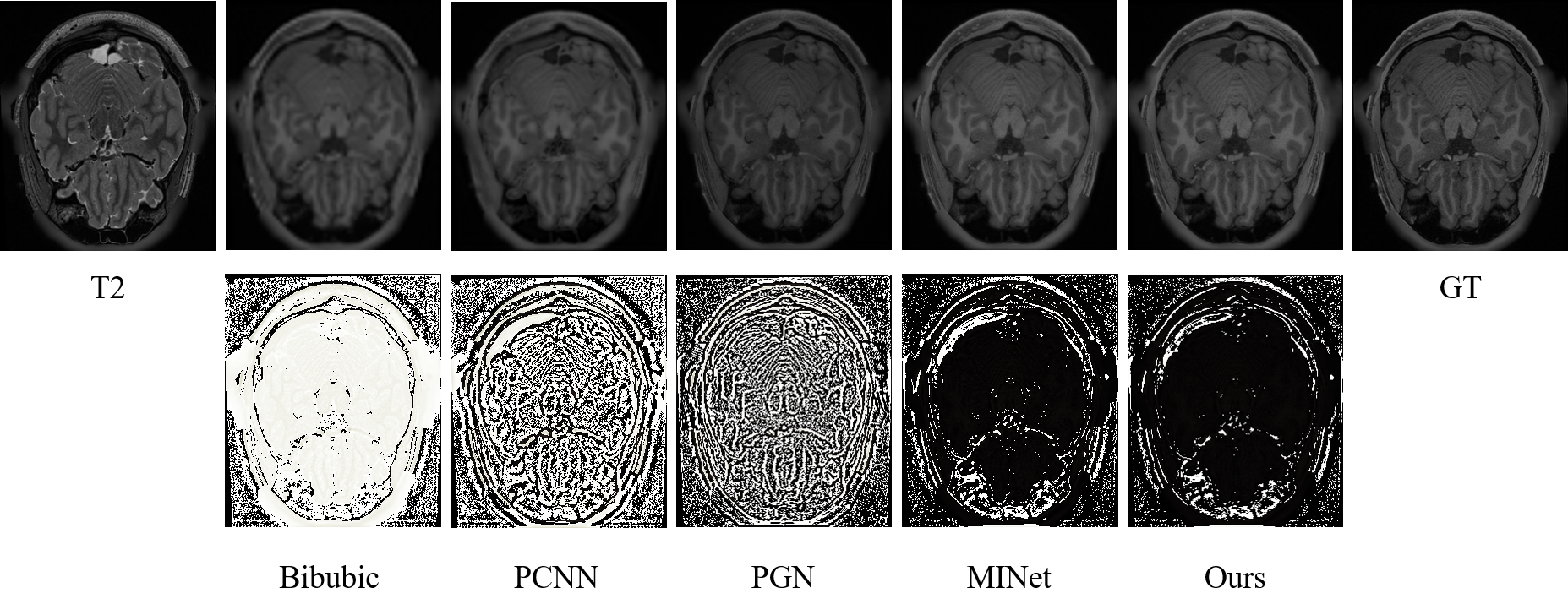}
\caption{Visual comparison of $\times 4$ MRI super-resolution and T2 denotes the guidance image. Images in the last row visualizes the MSE between the generated results and the ground truth.}
\label{mrifigb}
\end{center}
\end{figure*}

\begin{table*}[ht] 
	\normalsize
	\renewcommand{\tabcolsep}{3pt} % adjust horizontal space
\renewcommand{\arraystretch}{1.2}
\caption{Average performance comparison on the GaoFen2 datasets as the stage number increases. The best performance is shown in \textbf{bold}.}
\centering
\begin{tabular}{c|cccccccccc|} 
	\hline
Stage Number (K)& PSNR$\uparrow$ & SSIM$\uparrow$&SAM$\downarrow$&ERGAS$\downarrow$ &SCC$\uparrow$ &Q $\uparrow$ & $D_{\lambda}$ $\downarrow$& $D_{S}$ $\downarrow$ & QNR $\uparrow$\\ \hline
1 &  41.2772 & 0.9653  &    0.0249  &  1.0114   & 0.9664  &  0.7556 & 0.0616 & 0.1145	& 0.8319\\
2  & 41.4274 &	0.9673 &	0.0242  &	0.9834  &	0.9696 & 0.7650	& 0.0595 &	0.1106 &	0.8375\\ 
3  & 41.8058 &	0.9697 &	0.0224  &	0.9306  &  0.9737 &	0.7698 &	0.0622 &	0.1128 &	0.8329\\ 
4  & \textbf{41.8577}	& \textbf{0.9697}	 & \textbf{0.0229}	& \textbf{0.9420}	 & \textbf{0.9745}	& \textbf{0.7740}	 & \textbf{0.0629}	 & \textbf{0.1154}	& \textbf{0.8299}\\
5  & 41.7545 &	0.9690 &	0.0226 &	0.9431 &	0.9729 &	0.7699 &	0.0600 &	0.1166 &	0.8315\\
6 &41.4274 & 0.9673 &	0.0242 &	0.9834 &	0.9696 &	0.7650 &	0.0595 &	0.1106 &	0.8375 \\
	\hline
\end{tabular}
\label{tab:stage}
\end{table*}

\begin{table*}[t]
	\normalsize
	\renewcommand{\tabcolsep}{3pt} % adjust horizontal space
\renewcommand{\arraystretch}{1.2}
\caption{Experimental results on evaluating the effect of persistent memory module and non-local cross-modalities module. The best performance is shown in \textbf{bold}.}
\centering
\begin{tabular}{ccc|cccccccccc|} 
	\hline
Configurations& memory&non-local& PSNR$\uparrow$ & SSIM$\uparrow$&SAM$\downarrow$&ERGAS$\downarrow$ &SCC$\uparrow$ &Q $\uparrow$ & $D_{\lambda}$ $\downarrow$& $D_{S}$ $\downarrow$ & QNR $\uparrow$\\ \hline
(I)  & \XSolid & \Checkmark & 41.6287 &  0.9683  &   0.0237  &    0.9653  &   0.9727    & 0.7673  &    0.0641 &    0.1154 &     0.8287\\ 
(II)  & \Checkmark & \XSolid  & 41.7665&	0.9697&	0.0233&	0.9437&	0.9742&	0.7731&	0.0636&	0.1168&	0.8279
\\ 
\hline
Ours& \Checkmark&\Checkmark& \textbf{41.8577} & \textbf{0.9697} & \textbf{0.0229} & \textbf{0.9420}  & \textbf{0.9745}  & \textbf{0.7740}  & \textbf{0.0629}    &  \textbf{0.1154} & \textbf{0.8299}\\
	\hline
\end{tabular}
 \label{abla}
\end{table*}

\begin{figure*}
\begin{center}
\includegraphics[width=\textwidth]{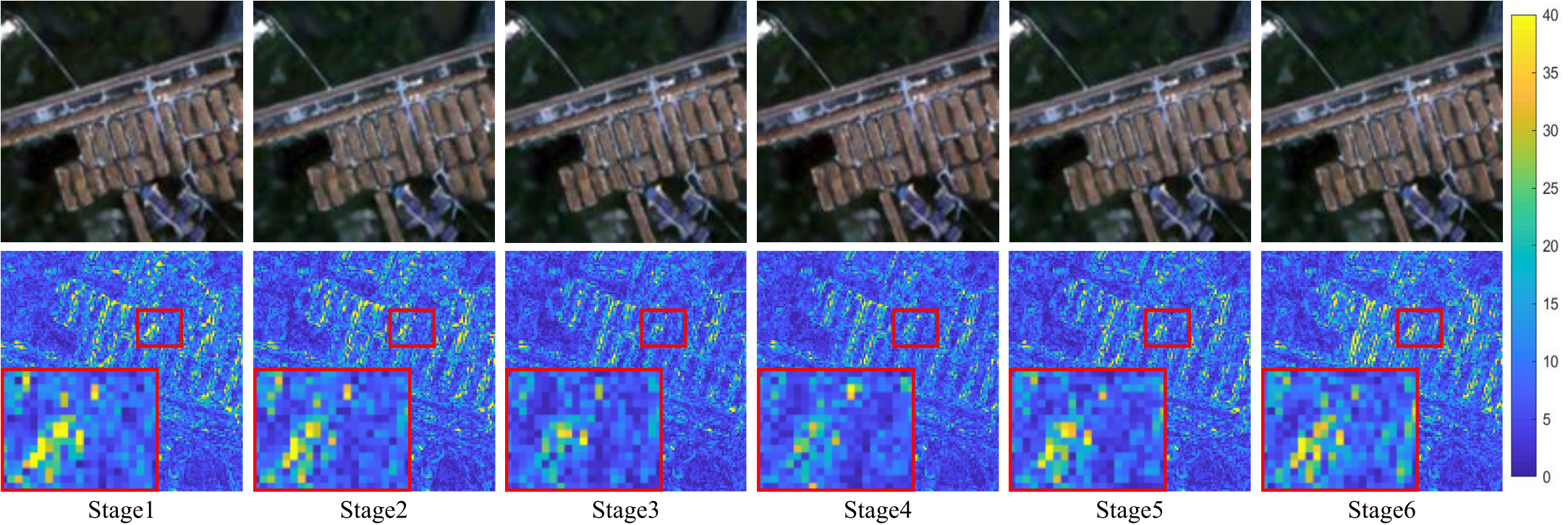}
\caption{Visual comparison of different stages.}
\label{stage}
\end{center}
\end{figure*}

\subsubsection{Compared with SOTA Methods}\label{exp:mridetail}
 The average quantitative results in $\times 4$ and $\times 2$ MR Image SR cases are reported in Table~\ref{tab-mri} and Table~\ref{tab-mri-2}, respectively. As can be seen, our proposed method outperforms all the other competing  methods in terms of all the indexes. Specifically, compared with the second best MINet~\citep{feng2021multi}, our method obtains an increase of 0.1 dB for PSNR, and 0.01 for SSIM in the $\times 4$ ratio case. Additionally, we also make visual comparisons on some representative samples in Fig.~\ref{mrifig} and Fig.~\ref{mrifigb}, where the first row illustrates the MR Image SR results and the second row visualizes the MSE residual between the generated HR MR images and the ground truth HR MR images. From Fig.~\ref{mrifig} and Fig.~\ref{mrifigb}, we can also observe that the generated HR MR images of our method have the best visual effect, and the corresponding MSE residual maps contain fewest structure information, which further supports the visual effect superiority of our method.

\subsection{Ablation Study}
To investigate the contribution of the devised modules in our proposed network, we have conducted comprehensive ablation studies on the Gaofen2 satellite dataset of the Pan-sharpening task. To be specific, the non-local cross-modalities module and persistent memory module are the two core designs. In addition, our proposed method is developed in the iterative unfolding manner. Therefore, the studies with respect to the number of stages and the parameter sharing mechanism across stages are also conducted. Furthermore, deepening into the memory module, we also explore the effect of different-location information transmission. All the experimental results are measured by the widely-used IQA metrics, i.e.,  ERGAS~\citep{ergas}, PSNR, SSIM, SCC, Q index, SAM~\citep{sam}, $D_\lambda$, $D_S$, and QNR.  

\vspace{0.8em}
\textbf{Effect of the stage number.} The effect of stage number is
first analyzed. The stage number T is set as $\{1,2,3,4,5,6\}$. The average quantitative results in terms of 8 metrics are reported in Table \ref{tab:stage}. Taking $K=1$ as the baseline, we can see that the network performance has a continuous improvement when the stage number $K$ increases to 4. This is because our proposed network has a more powerful feature extraction ability as the stage number increases. However, when the stage number is over 4, the quantitative results show a slight decreasing trend, which may be caused by the over-fitting problem due to the increasing stage. Therefore, we set $K=4$ as the default stage number in all the experiments to balance the performance and computational complexity. Additionally, to further undertake the effect of stage number $K$, we present the representative sample generated by different model variants with stage number from $K=1$ to $K=6$ in Fig.~\ref{stage}. From Fig.~\ref{stage}, we can also observe that the model with stage number $K=4$ obtains the best visual effect, which is consistent with the quantitative results of Table~\ref{tab:stage}.

\vspace{0.8em}
\textbf{Effect of parameter sharing.} In this experiment, the effect of the parameters sharing mechanism across stages is verified. We take the proposed network with 4 stages as the baseline to conduct the comparison. As well recognized, the proposed network has the same network architecture in each stage. The corresponding quantitative results are reported in Table \ref{tab:sharing}, from which we can observe that disabling parameter sharing is capable of improving the model performance to some extent. This may attribute to the increase of the model complexity when the parameter sharing mechanism is disabled. However, the increased model complexity will make the training of the model be slower. To achieve a good trade-off between the model parameter and model performance, we finally choose the model with parameter sharing as the default setting in all the experiments.

\begin{table}[ht]
\centering
\caption{
Performance comparison of parameter sharing mechanism on the GaoFen2 datasets. The best performance is shown in \textbf{bold}.}
\label{tab:sharing}
\renewcommand\arraystretch{1.2}
\begin{tabular}{c|cccc}
\toprule
{Parameter-sharing}          & PSNR $\uparrow$            & {SSIM}  $\uparrow$          & {SAM} $\downarrow$            & {ERGAS} $\downarrow$          \\ \midrule
With       & 41.8577         & 0.9697         & 0.0229          & 0.9420          \\
Without      & \textbf{42.1512} & \textbf{0.9724} & \textbf{0.0214} & \textbf{0.9042} \\ \bottomrule
\end{tabular}%
\vspace{-0.8em}
\end{table}

\vspace{0.8em}
\noindent\textbf{Effect of persistent memory module.} To explore the effectiveness of the memory module, we choose the model with memory module as the baseline and then obtain a variant of this model by eliminating the memory module from the baseline model. The quantitative results are reported in Table \ref{abla}, from which it can be seen that by eliminating the memory module from the baseline, the performance in terms of all the criteria decreases. This is because memory module can reduce the information loss from the feature channel transformation and facilitate the information interaction across stages. Therefore, we can conclude that the memory module is really helpful to improve the performance.

\vspace{0.8em}
\noindent\textbf{Effect of non-local cross-modalities module.} In the section, to verify the effectiveness of the non-local cross-modalities module, we choose the model with non-local cross-modalities module as the baseline and then obtain a variant of this model by eliminating this module from the baseline model. The quantitative results are reported in Table \ref{abla}. From the third and fourth row of Table~\ref{abla}, we can observe that by deleting the non-local cross-modalities module from the baseline, the performance with respect to all the metrics decreases. The reason is that the non-local cross-modality module is devised to enhance the spatial resolution of the LR target image by transferring the semantically-related and structure-consistent  content from the HR guidance image into the LR target image. Therefore, we can conclude that the non-local cross-modalities module is beneficial to our proposed model.

\begin{table}[ht]
\centering
\caption{
% \textbf{The effect of stage.} 
Performance comparison of different-location information on the GaoFen2 datasets. The best performance is shown in \textbf{bold}.}
\label{tab:location}
\renewcommand\arraystretch{1.2}
\begin{tabular}{c|cccc}
\toprule
{Location}          & PSNR $\uparrow$            & {SSIM}  $\uparrow$          & {SAM} $\downarrow$            & {ERGAS} $\downarrow$          \\ \midrule
Single       & 41.7199         & 0.9688         & 0.0235          & 0.9461          \\

Multiple      &  \textbf{41.8577} & \textbf{0.9697} & \textbf{0.0229} &  \textbf{0.9420} \\ \bottomrule
\end{tabular}%
\vspace{-0.8em}
\end{table}

\vspace{0.8em}
\noindent\textbf{Effect of memorizing the different-location information.} As shown in Fig.~\ref{sonfig}, we add the different-layer information into the memory module at three locations. Taking the ${\rm UNet}$ at $k$-iteration as an example, $\mathbf{HP}_1^{(k)}$, $\mathbf{HP}_2^{(k)}$ and $\mathbf{U}^{(k)}$ are adopted in the output and feature spaces. In this experiment, we regard this model as baseline (denoted as Multiple), and then obtain a variant of this model (denoted as Single) by only transmitting $\mathbf{U}^{(k)}$ into the memory module. The corresponding quantitative comparison results are reported in Table \ref{tab:location}. As can be seen, adding the memorized information at different locations will improve the performance of single location model to some extent. This is because more feature information is memorized and transformed into next iteration, and thus the information loss can be obviously alleviated. Therefore, we adopt the design of memorizing different-location information in our network. Similarly, ${\rm VNet}$ and ${\rm HNet}$ have similar implementations of memorizing the different-location information.

\section{Conclusion}
\vspace{-.2cm}
In this work, we first propose a maximal a posterior (MAP) estimation model for guided image super-resolution (GISR) by embedding the proposed global implicit and explicit priors on the HR target image. We then design a novel alternating minimization algorithm to solve this model, and then unfold the proposed algorithm into a deep network. To facilitate the signal flow across unfolding stages, the persistent memory mechanism is introduced to augment the information representation by exploiting the Long short-term memory unit (LSTM) in the image and feature spaces. In this way, both the interpretability and representation ability of the deep network are improved. Extensive experiments validate the superiority of our method on three representative GISR tasks, including Pan-sharpening, Depth Image SR, and MR Image SR.

% \begin{acknowledgements}
%   This work was supported in part by Samsung Research Funding \& Incubation Center for Future Technology (SRFC-IT1802-06), the Louis Vuitton/ENS chair on artificial intelligence, the Inria/NYU collaboration agreement, and the French government under management of Agence Nationale de la Recherche as part of the ``Investissements d'avenir" program, reference ANR-19-P3IA-0001 (PRAIRIE 3IA Institute).
% \end{acknowledgements}


% TODO: bib style need to be changed
\bibliographystyle{spbasic}      % basic style, author-year citations
\bibliography{egbib} % name your BibTeX data base

\end{document}